\documentclass[sigconf]{acmart} 

\usepackage{tikz}
\usetikzlibrary{positioning}
\usepackage{booktabs}
\usepackage{longtable}
\usepackage{array}
\usepackage{xcolor}
\usepackage{soul}
\usepackage{pifont}
\usepackage{makecell}
\usepackage{rotating}     % sidewaysfigure -- the quadrant
\usepackage{pdflscape}    % landscape -- only the held-back reasons appendix uses it
\usepackage{colortbl}

\usepackage{enumitem}
\usepackage{hyperref}
\definecolor{GCblue}{HTML}{1F6FB2}   % Ability
\definecolor{GCgreen}{HTML}{177A2E}  % Cooperativeness
\definecolor{GCsand}{HTML}{A97A00}   % Integrity
\definecolor{GCrose}{HTML}{C0341A}   % Benevolence
\definecolor{GCahead}{HTML}{3D4C5E}  % Getting ahead (motive ring)
\definecolor{GCalong}{HTML}{6E4A52}  % Getting along (motive ring)
\definecolor{GCtodo}{HTML}{A03030}

\newcommand{\site}{Microsoft}

\newif\ifshowsupp
\showsuppfalse
\definecolor{GCquote}{HTML}{1F4E5A}
\newcommand{\q}[1]{{\color{GCquote}``#1''}}

\newcommand{\suitmark}[2]{\,\textcolor{#1}{\raisebox{0.05ex}{\small #2}}}
\newcommand{\markAb}{\suitmark{GCblue}{\ding{171}}}   % spade
\newcommand{\markCo}{\suitmark{GCgreen}{\ding{168}}}  % club
\newcommand{\markIn}{\suitmark{GCsand}{\ding{169}}}   % diamond
\newcommand{\markBe}{\suitmark{GCrose}{\ding{170}}}   % heart

\newcommand{\Ab}[1]{\emph{#1}\markAb}
\newcommand{\Co}[1]{\emph{#1}\markCo}
\newcommand{\In}[1]{\emph{#1}\markIn}
\newcommand{\Be}[1]{\emph{#1}\markBe}

\newcommand{\bAb}{\textbf{Ability}\markAb}
\newcommand{\bCo}{\textbf{Cooperativeness}\markCo}
\newcommand{\bIn}{\textbf{Integrity}\markIn}
\newcommand{\bBe}{\textbf{Benevolence}\markBe}

\definecolor{GCdetr}{HTML}{D9D9D9}
\definecolor{GCnot}{HTML}{BDBDBD}
\definecolor{GCsome}{HTML}{8FA9C4}
\definecolor{GCvery}{HTML}{4C7BAF}
\definecolor{GCess}{HTML}{1B3F6B}

\newcommand{\arch}[1]{\emph{#1}}

\usepackage[most]{tcolorbox}
\newtcolorbox{takeaway}[1][]{
  enhanced, breakable, boxrule=0pt, frame hidden,
  colback=black!4, borderline west={2.2pt}{0pt}{black!45},
  left=7pt, right=7pt, top=5pt, bottom=5pt,
  before skip=7pt, after skip=9pt, #1}
\newcommand{\tkw}[1]{\textbf{\textsf{\small #1}}:\hspace{1mm}}

\setcopyright{none}
\renewcommand\footnotetextcopyrightpermission[1]{}   % removes the copyright/conference footnote

\newcommand{\figwheel}{%
\begin{figure*}[t]\centering
\includegraphics[width=\textwidth]{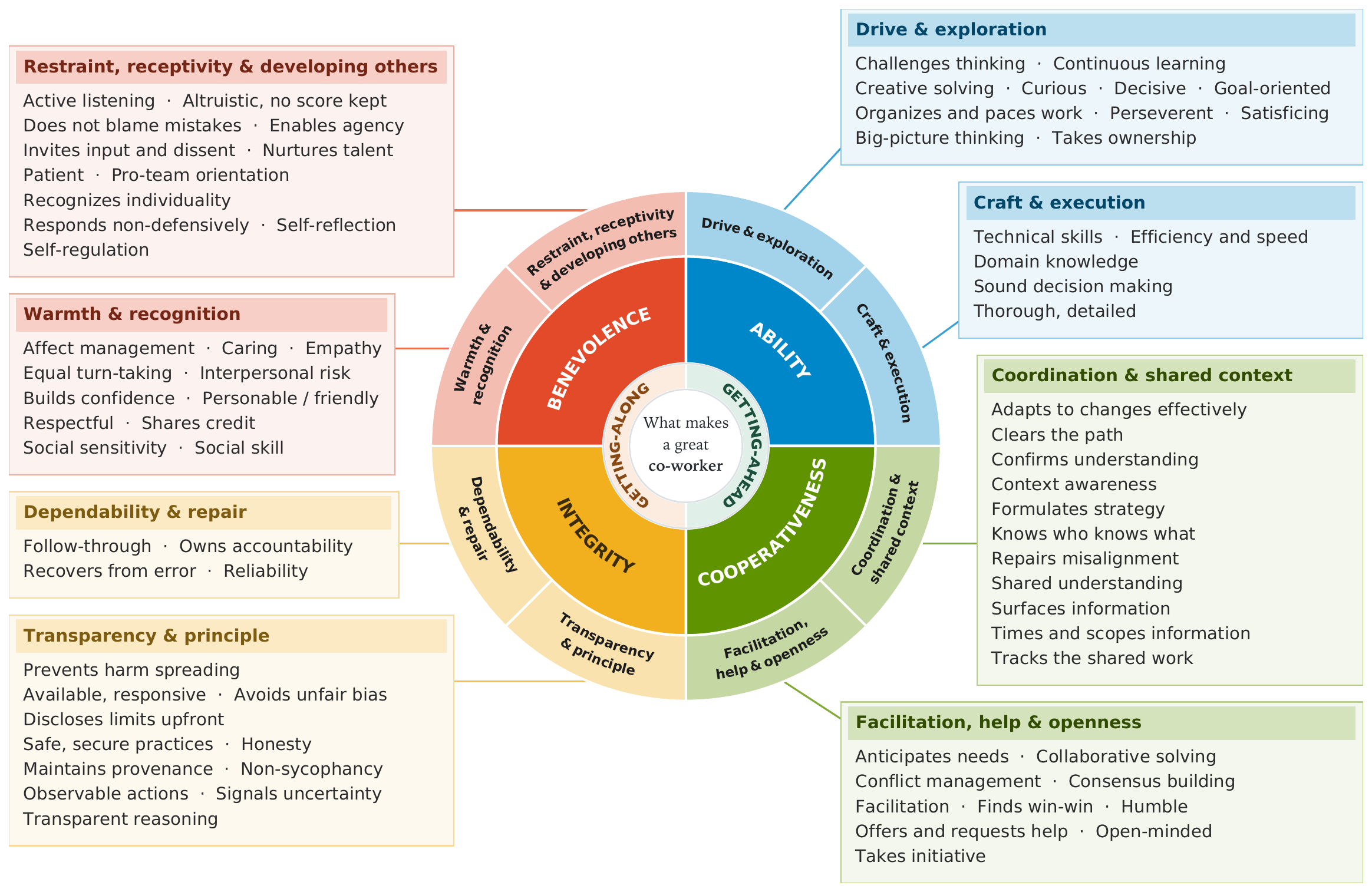}
\caption{\small The BACI framework: 75 qualities of a great co-worker organized into four buckets under the two broad motives of working life: \bAb{} and \bCo{} reflect \emph{getting ahead}, \bIn{} and \bBe{} reflect \emph{getting along}. From the center outward, the wheel shows these two motives, four buckets, and eight subdimensions; the surrounding panels list their constituent qualities.}
\Description{A circular diagram of the BACI framework, with three concentric rings surrounded by eight rectangular panels arranged in two stacks of four. The center asks, ``What makes a great co-worker?'' The innermost ring divides the wheel into two motives: getting along on the left and getting ahead on the right. The middle ring contains four color-coded quadrants: Benevolence in red at the upper left, Ability in blue at the upper right, Cooperativeness in green at the lower right, and Integrity in yellow at the lower left. Benevolence and Integrity align with getting along; Ability and Cooperativeness align with getting ahead. The outer ring divides each quadrant into two subdimensions, each connected by a matching colored line to a panel listing its qualities. The four panels on the left are, from top to bottom: Restraint, receptivity and developing others, listing active listening; altruism with no score kept; not blaming mistakes; enabling agency; inviting input and dissent; nurturing talent; patience; pro-team orientation; recognizing individuality; responding non-defensively; self-reflection; and self-regulation. Warmth and recognition lists affect management; caring; empathy; equal turn-taking; interpersonal risk; building confidence; being personable or friendly; respectfulness; sharing credit; social sensitivity; and social skill. Dependability and repair lists follow-through; owning accountability; recovering from error; and reliability. Transparency and principle lists preventing harm from spreading; being available and responsive; avoiding unfair bias; disclosing limits upfront; safe and secure practices; honesty; maintaining provenance; non-sycophancy; observable actions; signaling uncertainty; and transparent reasoning. The four panels on the right are, from top to bottom: Drive and exploration, listing challenging thinking; continuous learning; creative solving; curiosity; decisiveness; being goal-oriented; organizing and pacing work; perseverance; satisficing; big-picture thinking; and taking ownership. Craft and execution lists technical skills; efficiency and speed; domain knowledge; sound decision-making; and thorough, detailed work. Coordination and shared context lists adapting to changes effectively; clearing the path; confirming understanding; context awareness; formulating strategy; knowing who knows what; repairing misalignment; shared understanding; surfacing information; timing and spacing information; and tracking shared work. Facilitation, help and openness lists anticipating needs; collaborative solving; conflict management; consensus building; facilitation; finding win-win solutions; humility; offering and requesting help; open-mindedness; and taking initiative.}
\label{fig:wheel}\end{figure*}}

\newcommand{\figranking}{%
\begin{figure*}[h]\centering
\includegraphics[width=\textwidth]{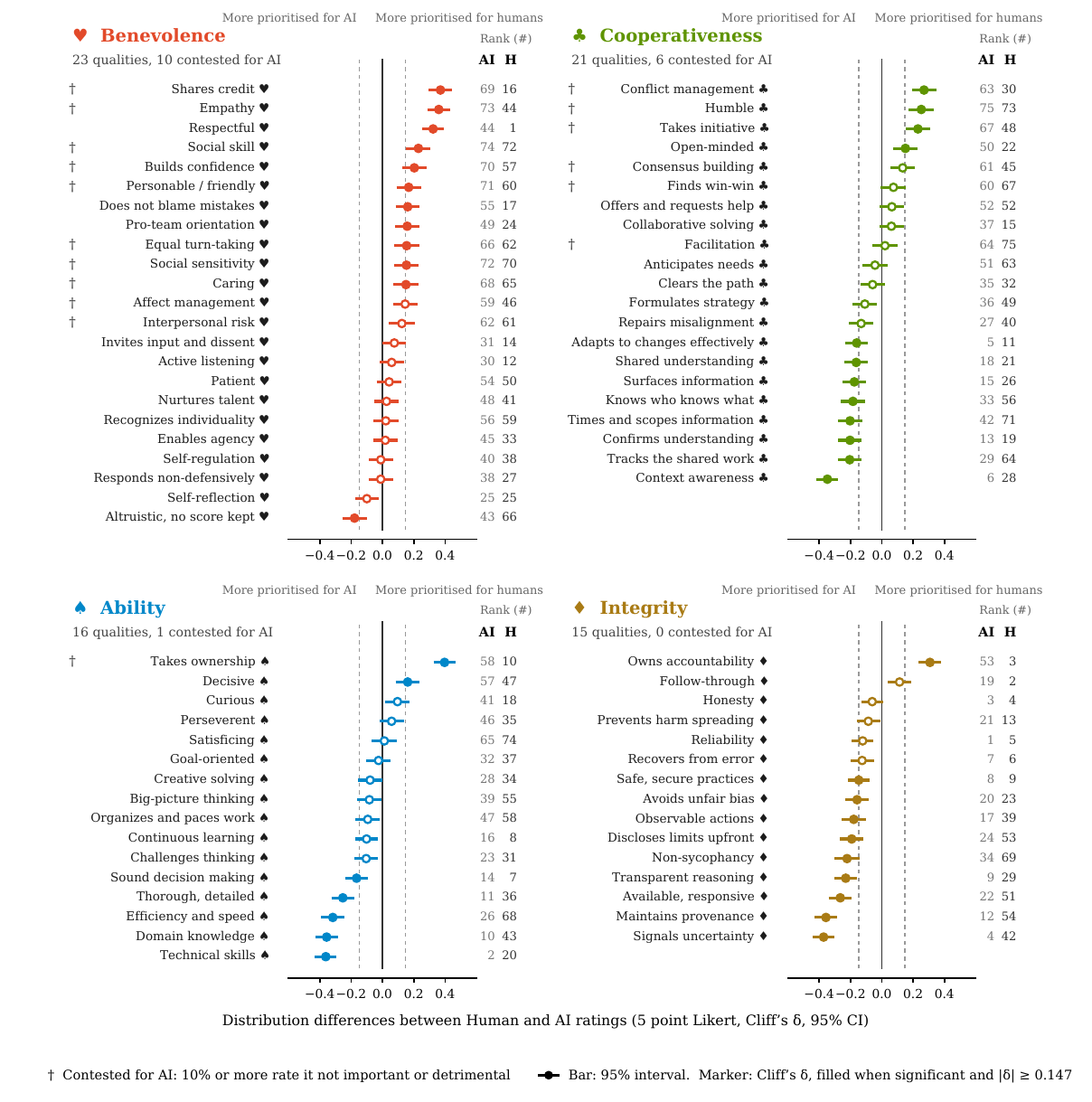}
\caption{\small Human--AI differences in priorities for co-worker qualities, grouped by BACI bucket. Markers show Cliff's $\delta$ comparing the Likert distributions; bars show 95\% confidence intervals. Positive values indicate higher priority for humans; negative values indicate higher priority for AI. Filled markers indicate differences meeting both the corrected significance criterion and the $|\delta|$ threshold. A dagger ($\dagger$) marks contested qualities that AI-referent respondents are divided on. Columns labeled \emph{AI} and \emph{H} are a quality's rank out of 75 within each referent.}
\Description{Four forest plots arranged in a two-by-two grid. The upper-left panel shows Benevolence in red with a heart symbol: 23 qualities, 10 contested for AI. The upper-right panel shows Cooperativeness in green with a club symbol: 21 qualities, six contested for AI. The lower-left panel shows Ability in blue with a spade symbol: 16 qualities, one contested for AI. The lower-right panel shows Integrity in gold with a diamond symbol: 15 qualities, none contested for AI. Each panel contains one row per quality, ordered from the largest human-favoring difference at the top to the largest AI-favoring difference at the bottom. Quality names appear on the left, point estimates and horizontal 95 percent confidence intervals in the center, and two rank columns labeled AI and H on the right, giving each quality's rank among all 75 within each referent. The horizontal axis shows Cliff's delta: negative values favor AI and positive values favor humans. A solid vertical line marks zero; dashed lines mark the substantive thresholds at minus and plus 0.147. Filled markers meet both the corrected significance criterion and the substantive threshold; hollow markers do not meet both. Daggers beside quality names identify those rated not important or detrimental by at least 10 percent of AI-referent respondents.}
\label{fig:ranking}\end{figure*}}

\newcommand{\figquadrant}{%
\begin{figure*}[t]\centering
\includegraphics[width=0.95\textwidth]{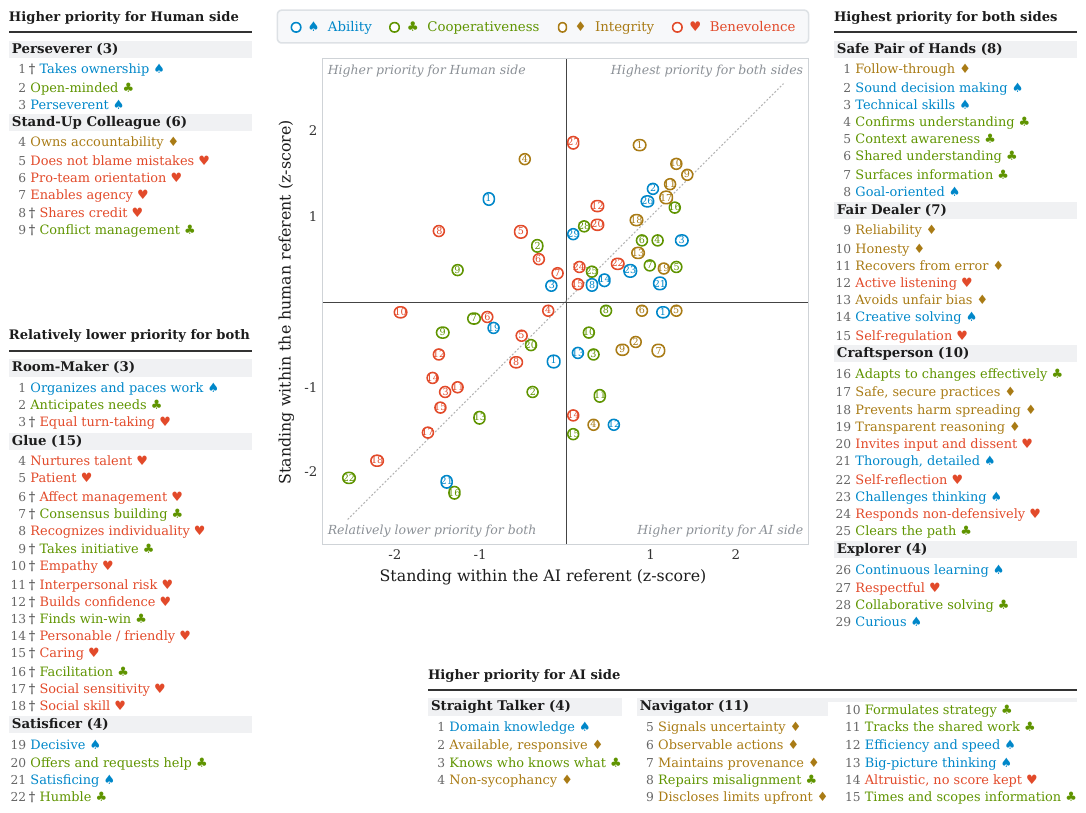}
\caption{Relative priorities for 75 co-worker qualities and their grouping into 11 co-worker archetypes. Each point in the z-score plot represents a quality's standardized priority within the AI referent (x-axis) and human referent (y-axis). The axes divide qualities into four priority regions; the dashed diagonal indicates equal relative standing across referents. Surrounding lists group qualities into archetypes within each region. Point numbers match the numbered qualities and restart in each region; parentheses give the number of qualities per archetype. Colors and suit symbols identify BACI buckets, and daggers mark qualities contested for AI.}
\Description{A central scatterplot surrounded by four sets of archetype lists. The horizontal axis shows standardized priority within the AI referent, and the vertical axis shows standardized priority within the human referent. Solid lines at zero divide the plot into four quadrants; a dashed diagonal from bottom left to top right marks equal relative standing. Each numbered point represents one quality, colored by BACI bucket: blue for Ability, green for Cooperativeness, gold for Integrity, and red for Benevolence. A legend above the plot pairs these colors with spade, club, diamond, and heart symbols. The upper-left quadrant represents higher relative priority for humans only. It corresponds to a list at the upper left containing Perseverer, with three qualities, and Stand-Up Colleague, with six. The lower-left quadrant represents relatively lower priority for both referents and corresponds to a list below it containing Room-Maker, with three qualities; Glue, with 15; and Satisficer, with four. The upper-right quadrant represents higher relative priority for both referents and corresponds to a tall list on the right containing Safe Pair of Hands, with eight qualities; Fair Dealer, with seven; Craftsperson, with ten; and Explorer, with four. The lower-right quadrant represents higher relative priority for AI only and corresponds to a list below the plot containing Straight Talker, with four qualities, and Navigator, with 11. Each list has shaded archetype headings followed by numbered quality names and their bucket symbols. Numbers restart within each priority region and link list entries to plotted points. Daggers identify qualities contested for AI.}
\label{fig:quadrant}\end{figure*}}

\begin{document}

\title{What Makes a Great Co-Worker in an AI-Native Workplace?}

% \author{%
%   Rudrajit Choudhuri$^{1}$\quad 
%   Max Meijer$^{2}$ \quad
%   Sam Yu-Te Lee$^{3}$ \quad
%   Cinoo Lee$^{2}$ \quad
%   Caolan Mannion$^{2}$ \quad\\
%   Peter Jahn$^{2}$ \quad
%   Anita Sarma$^{1}$ \quad
%   Christian Bird$^{2}$ \quad
%   Alice Ferng$^{2}$
% }
% \affiliation{%
%   $^{1}$Oregon State University, OR, USA. Email: \{choudhru, sarmaa\}@oregonstate.edu \\
%   $^{2}$Microsoft, USA. Email: \{maxmeijer, cinoolee, caolanm, pejahn, cbird, alferng\}@microsoft.com \\
%   $^{3}$University of California Davis, CA, USA. Email: ytlee@ucdavis.edu%
% }

\author{%
  Rudrajit Choudhuri$^{1}$ \quad
  Max Meijer$^{2}$ \quad
  Sam Yu-Te Lee$^{3}$ \quad
  Cinoo Lee$^{2}$ \quad
  Caolan Mannion$^{2}$ \quad\\
  Peter Jahn$^{2}$ \quad
  Anita Sarma$^{1}$ \quad
  Christian Bird$^{2}$ \quad
  Alice Ferng$^{2}$
}
\affiliation{%
  $^{1}$Oregon State University, OR, USA. Email: \{choudhru, sarmaa\}@oregonstate.edu \\
  $^{2}$Microsoft, USA. Email: \{maxmeijer, cinoolee, caolanm, pejahn, cbird, alferng\}@microsoft.com \\
  $^{3}$University of California Davis, CA, USA. Email: ytlee@ucdavis.edu%
}
\renewcommand{\shortauthors}{Choudhuri et al.}

%  DONT DELETE THIS: AI as a coworker: You work in a continuous interdependent back and forth with AI on a given knowledge work task, jointly shaping outputs and decisions. 

% Shared oversight: you and the AI share responsibility over the workflow about the role. You handle some tasks and decisions independently, while the AI does the same on others. You stay informed on what the AI is doing but do not need to approve every action. Control shifts between you and the AI depending on the task. 

% Collaborative knowledge work is changing in ways LLM agents are now embedded in how teams research, design, write, and decide: mediating between members, synthesizing inputs, reformulating ideas, and drafting shared outputs. They do not only facilitate collaboration; they operate within the workflow at the moment contributions are being formed. In doing so, they risk undermining the social conditions under which contributions can be witnessed, attributed, and held accountable. 

% Use this paper for the intro: https://arxiv.org/pdf/2607.26387 

\begin{abstract}
As knowledge work grows interdependent between humans and AI, we ask what makes a great co-worker in an AI-native workplace. To answer this, we conducted 22 interviews and a large-scale mixed-methods survey of 1,534 knowledge workers at a multinational technology company. We contribute BACI, a framework of 75 co-worker qualities that apply to humans and AI, spanning \textit{Benevolence}, \textit{Ability}, \textit{Cooperativeness}, and \textit{Integrity}. Comparing priorities for humans and AI identified 11 co-worker archetypes and revealed disagreement over whether AI should have warmth, take initiative, or own outcomes. We also show how priorities for these archetypes varied with workers' individual characteristics. Lastly, we contribute a taxonomy of \textit{AI work etiquette} capturing the obligations co-workers expect of one another when preparing, sharing, and taking responsibility for AI-supported work. Based on these findings, we derive implications to inform worker-centric AI and workplace design.

\end{abstract}

% \begin{CCSXML}
% <ccs2012>
%    <concept>
%        <concept_id>10003120.10003121.10003126</concept_id>
%        <concept_desc>Human-centered computing~HCI theory, concepts and models</concept_desc>
%        <concept_significance>500</concept_significance>
%        </concept>
%    <concept>
%        <concept_id>10003120.10003121.10011748</concept_id>
%        <concept_desc>Human-centered computing~Empirical studies in HCI</concept_desc>
%        <concept_significance>500</concept_significance>
%        </concept>
%     <concept>
%        <concept_id>10003120.10003130</concept_id>
%        <concept_desc>Human-centered computing~Collaborative and social computing</concept_desc>
%        <concept_significance>300</concept_significance>
%        </concept>
%    <concept>
%        <concept_id>10003120.10003130.10003131.10003570</concept_id>
%        <concept_desc>Human-centered computing~Computer supported cooperative work</concept_desc>
%        <concept_significance>300</concept_significance>
%        </concept>
%  </ccs2012>
% \end{CCSXML}

% \ccsdesc[500]{Human-centered computing~HCI theory, concepts and models}
% \ccsdesc[500]{Human-centered computing~Empirical studies in HCI}
% \ccsdesc[300]{Human-centered computing~Collaborative and social computing}
% \ccsdesc[300]{Human-centered computing~Computer supported cooperative work}


% \begin{teaserfigure}
%     \centering
%     \textcolor{red}{\fbox{**Teaser figure**}}  
%     % \includegraphics[width=.1\linewidth]{figure/teaser.png}
%     \caption{**Our teaser figure here (\textcolor{blue}{Have a look into the example: https://dl.acm.org/doi/pdf/10.1145/3708359.3712086})**.
%     }
%     \Description{Enjoying the baseball game from the third-base
%     seats. Ichiro Suzuki preparing to bat.}
%     \label{fig:teaser}
%  \end{teaserfigure}

\begin{teaserfigure}\centering
\includegraphics[width=0.89\textwidth]{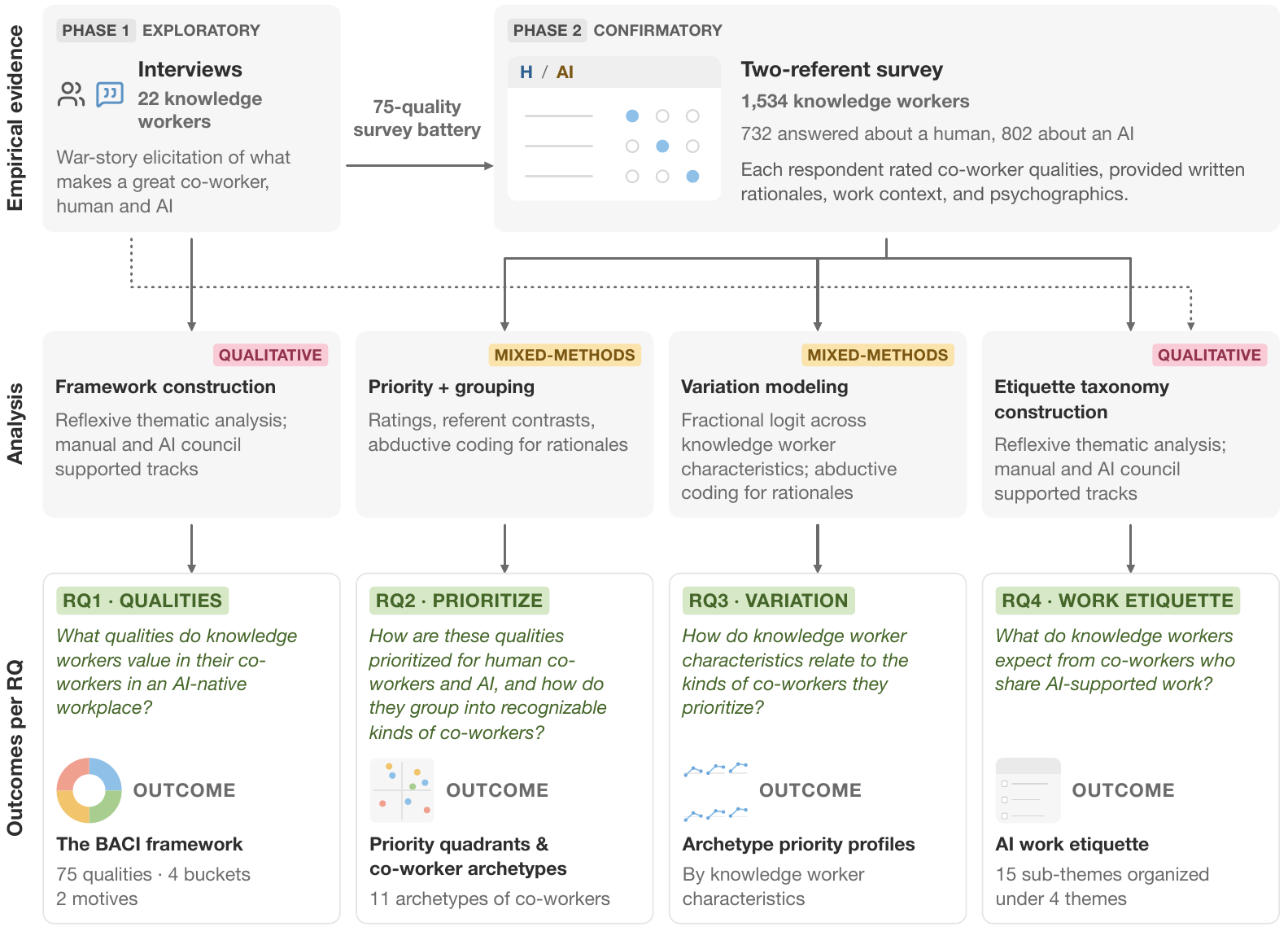}
\caption{Research overview of our two-phase mixed-methods study. In the exploratory phase, interviews with 22 knowledge workers produced the BACI framework of 75 co-worker qualities (RQ1). In the confirmatory phase, a survey of 1,534 knowledge workers established how these qualities were prioritized in human coworkers and AI, identified 11 co-worker archetypes (RQ2), and revealed how archetype priorities varied across workers' individual characteristics (RQ3). Both phases informed a taxonomy of AI work etiquette describing what co-workers expect of one another when sharing AI-supported work (RQ4).}
\Description{A flow diagram arranged in three horizontal rows labeled Empirical evidence, Analysis, and Outcomes per RQ. The middle and bottom rows each contain four panels aligned vertically into four columns. The top row contains two panels: Phase 1, exploratory, above the first column, and a wider Phase 2, confirmatory, spanning the remaining three columns. Phase 1 describes war-story interviews with 22 knowledge workers about great human and AI co-workers. A rightward arrow labeled ``75-quality survey battery'' connects it to Phase 2, a two-referent survey of 1,534 knowledge workers: 732 answered about a human and 802 about an AI. Respondents rated co-worker qualities and provided written rationales, work context, and psychographics. In the middle row, the four analysis panels are, from left to right: qualitative framework construction using reflexive thematic analysis with manual and AI council-supported tracks; mixed-methods priority and grouping using ratings, referent contrasts, and abductive coding of rationales; mixed-methods variation modeling using fractional logit across worker characteristics and abductive coding of rationales; and qualitative etiquette taxonomy construction using reflexive thematic analysis with manual and AI council-supported tracks. A downward arrow connects the interviews to framework construction, and a dotted connector links them to etiquette taxonomy construction. Survey arrows branch downward to the other three analysis panels. Each analysis panel connects downward to its corresponding research-question and outcome panel. From left to right, these are: RQ1, Qualities, yielding the BACI framework with 75 qualities, four buckets, and two motives; RQ2, Prioritize, yielding priority quadrants and 11 co-worker archetypes; RQ3, Variation, yielding archetype-priority profiles by worker characteristics; and RQ4, Work etiquette, yielding 15 sub-themes organized under four themes.}
\label{fig:method}
\end{teaserfigure}

\makeatletter
\let\ACM@savedClassError\ClassError
\def\ClassError#1#2#3{\ClassWarning{#1}{#2}}
\makeatother

\maketitle

\makeatletter
\let\ClassError\ACM@savedClassError
\makeatother

% THE POINT OF THE INTRODUCTION IS TO MAKE READERS WANT TO READ THE PAPER. DONT GIVE THEM RESULTS HERE. THE POINT OF THE CONCLUSION IS TO MAKE READERS FEEL GLAD THEY READ THE PAPER. WRITE 3 main results and a sunset there. 

\section{Introduction}
\label{sec:intro}

Knowledge workers have always divided problems, combined expertise, and built on each other’s outputs~\cite{thompson2017organizations,vandervegt2001patterns,faraj2000coordinating}. Depending on someone naturally leads to assessing them, and as a group-living species we have always been good at it; our standing and safety turn on \textit{getting along} and \textit{getting ahead} with people~\citep{hogan2013socioanalytic}. Ask someone about a great co-worker they have had, and they will most likely describe someone good at the work and good to work with. Technical skill matters, but so do coordination, candor, reliability, care, and the ability to make expertise usable by others. Decades of studies have confirmed and refined this folk knowledge~\citep{li2020distinguishes, kalliamvakou2019manager, dias2021maintainer, oleary2012skillset}, and theories of teamwork~\citep{salas2005bigfive,marks2001temporally}, organizational trust~\citep{mayer1995integrative}, and psychological safety~\citep{edmondson1999psychological} have explained why.

In the last couple of years, generative AI tools (hereafter AI tools) have entered this chain of dependence. As of 2026, 42\% of U.S. employees in the technology sector reported using AI daily in their role~\citep{gallup2026indicator}. Workers use these tools for a wide variety of cognitive tasks, including research, design, writing, analysis, problem-solving, and decision-making~\citep{brachman2024knowledge,lee2025impact,microsoft2026wti,yu2026adoption}. Because these tools communicate conversationally, people also often tend to respond to them socially~\citep{nass2000machines}, inviting framings such as AI ``teammates''~\citep{seeber2020machines,zhang2021ideal,weng2026friend}.

Accordingly, researchers have studied whether workers rely on AI appropriately~\citep{bucinca2021trust,bansal2021does,tankelevitch2024metacognitive,passi2024appropriate,choudhuri2025trust}, how they fit AI tools into their work~\citep{brachman2024knowledge,russo2024navigating,lee2025impact,choudhuri2026ai}, and what they expect of an AI teammate~\citep{zhang2021ideal,oneill2022humanautonomy,seeber2020machines}. More recently, work has begun to show that AI also reshapes teamwork~\citep{dellacqua2026cybernetic,yu2026adoption}, altering trust, attribution, and relationships among the people exchanging work~\citep{hwang2026forgiveness,hilbolling2026tensions,liebscher2026workslop}.

These changes become increasingly consequential as AI use grows routine across organizations and AI-supported work regularly enters shared workflows~\citep{yu2026adoption,msft_nfw,microsoft2026wti}. 
We call such workplaces \emph{AI-native}. We use the term \emph{\textbf{co-worker}} to refer to a practical role within such interdependent work. 
Sharing this role, however, does not necessarily make humans and AI equivalent. Although they may produce similar work, they may not bear the same obligations within the workplace. 
For example, humans can take ownership, be held accountable, and repair relationships; AI may not necessarily bear these obligations. 
Additionally, the process behind an AI's contribution is often opaque~\citep{tankelevitch2024metacognitive}, and its agreement, however fluent, may reflect how it was trained~\citep{sharma2024sycophancy,sarkar2024ai}. 
So the qualities workers prioritize in humans and AI have reason to differ. 
% AI may also change what workers value in one another: when fluent output is abundant, taste, discernment, and ownership of outcomes may become more consequential~\citep{lee2025impact,microsoft2026wti,choudhuri2026ai}.

The question, then, becomes which qualities matter for both referents, which are prioritized differently between them and why, and what remains a human obligation when AI-supported work is passed from one to another. In this paper, we thus investigate:

\begin{itemize}[label={},leftmargin=0pt]
\item \textbf{RQ1 --- Qualities.} What qualities do knowledge workers value in their co-workers in an AI-native workplace? (\S\ref{sec:framework})
\item \textbf{RQ2 --- Priorities.} How are these qualities prioritized for human co-workers and AI, and how do they group into recognizable kinds of co-workers? (\S\ref{sec:prevalence})
\item \textbf{RQ3 --- Variation.} How do knowledge worker characteristics relate to the kinds of co-workers they prioritize? (\S\ref{sec:whowantswhat})
\item \textbf{RQ4 --- Work etiquette.} What do knowledge workers expect from co-workers who share AI-supported work? (\S\ref{sec:emerging})
\end{itemize}

To answer these questions, we conducted a two-phase mixed-methods empirical study (Figure~\ref{fig:method}) at \site{}.
In the \textit{exploratory} phase, we interviewed 22 knowledge workers to identify the qualities they valued in their human co-workers and AI. We grounded our analysis in organizational psychology, teamwork science, and human--AI collaboration research~\citep{hogan2013socioanalytic,mayer1995integrative,seeber2020machines}. The interviews produced the \textbf{BACI framework}: 75 qualities of a great co-worker under \textit{Benevolence}, \textit{Ability}, \textit{Cooperativeness}, and \textit{Integrity}. In the \textit{confirmatory} phase, we surveyed 1,534 knowledge workers: 732 rated the qualities for a human co-worker and 802 for an AI. We phrased each quality identically for both referents, allowing comparison without attributing an inner state to either. The survey established how strongly each quality was prioritized or contested for humans and AI, grouped these qualities into 11 co-worker archetypes by their relative importance for both referents, and showed how priorities for those archetypes varied with workers' individual characteristics. Both phases also informed what workers expected of one another when a co-worker's AI-supported work reached them; we report these obligations as a taxonomy of \emph{AI work etiquette}.

% This paper makes four contributions:

% \begin{itemize}
% \item The \textbf{BACI framework}: 75 qualities of a great co-worker under two motives of working life, each askable of a person and of an AI.
% \item A comparative account of \textbf{how those qualities are prioritized}: which are wanted of both referents, which of one more than the other, and which are contested, grouped into 11 archetypes of great co-workers.
% \item An account of \textbf{how priorities vary with workers' individual characteristics}: work context, psychographics, and stance towards who counts as a co-worker and what they are for.
% \item A taxonomy of \textbf{AI work etiquette}: obligations workers expect of one another when preparing and sharing AI-supported work.
% \end{itemize}

Based on our work, we call on the HCI community to champion responsible work handoffs as a first-class concern of AI-native work, designing tools and workplaces as much for those who receive AI-supported work as for those who produce it (\S\ref{sec:discussion}).

\section{Related work}
\label{sec:related}

\subsection{Qualities of effective co-workers}
\label{sec:rel:human}

Studies across occupations show that effective contribution extends beyond technical performance~\citep{li2020distinguishes,kalliamvakou2019manager,dias2021maintainer,oleary2012skillset}. For example, great software engineers were distinguished by how they made decisions, communicated, and affected their teammates~\citep{li2020distinguishes}; managers by how they communicated, cleared obstacles, and nurtured talent~\citep{kalliamvakou2019manager}; and successful collaborators by their personal and interpersonal skills ahead of substantive expertise~\citep{oleary2012skillset}. 

Organizational research offers one explanation for why these qualities matter. Formal roles cannot fully specify every contribution an organization needs, so workers often contribute beyond their assigned roles by helping colleagues, maintaining the shared working environment, suggesting improvements, and raising concerns about risks or harm~\cite{organ1988ocb,liang2012voice}. Conversely, a co-worker may fulfill their assigned tasks while failing to sustain the context in which others can accomplish theirs, for example by withholding information, failing to raise concerns, or leaving coordination work undone. Whether workers make these broader contributions depends partly on felt obligation and psychological safety, which enable people to speak up and teams to learn from mistakes~\citep{liang2012voice,edmondson1999psychological}.

Teamwork research further explains that interdependent work demands  \emph{cooperativeness} across recurring episodes of goal-directed activity in which members monitor one another’s performance, maintain compatible understandings of roles and tasks, provide backup, and adapt as conditions change~\cite{marks2001temporally,salas2005bigfive,cannon1998individual,dechurch2010cognitive}. In these environments, psychological safety allows members to raise uncertainty, mistakes, and disagreement before they become failures~\citep{edmondson1999psychological}, while transactive knowledge allows them to coordinate effectively when they do not possess it themselves~\citep{faraj2000coordinating,woolley2010evidence}.

Moreover, interdependent work also requires accepting vulnerability to others. \citet{mayer1995integrative} identify \emph{ability}, \emph{benevolence}, and \emph{integrity} as the bases of perceived trustworthiness: whether another party can perform, intends well, and adheres to acceptable principles. Research on trust in automation and AI extends this concern to non-human agents and likewise shows that performance alone does not determine when reliance is warranted~\citep{lee2004trust,glikson2020human,kaplan2023trust,devisser2016almost}.

% In our study, we used \emph{cooperativeness} and the three bases of trustworthiness as the four buckets to organize our findings (\S\ref{sec:method}).

These accounts consistently combine two sides of effective contribution: accomplishing the work and sustaining the relationships that make collective work possible. Socio-analytic theory explains this recurring structure through two motives people pursue at work: \emph{getting ahead}, by attaining the status and resources associated with strong performance, and \emph{getting along}, by securing the acceptance and cooperation needed to work well with others~\citep{hogan2003socioanalytic}. Social-perception research finds the same two dimensions in how people judge one another, described as agency and communion~\citep{abele2007agency,fiske2007universal}. We therefore use these two motives as the higher-level structure within which we situate \emph{benevolence}, \emph{ability}, \emph{cooperativeness}, and \emph{integrity} in our framework.

Overall, this literature identifies qualities of effective co-workers in human teams. However, as AI use becomes routine in workplaces, it remains unclear which qualities workers value in human and AI contributors, where their expectations align or differ, and what obligations they owe each other when sharing AI-supported work. We address these questions in this study.

\subsection{AI as a contributor to interdependent work}
\label{sec:rel:ai}

Long before generative AI, automation research asked what a machine must do to participate in joint activity. \citet{klein2004ten} identified ten challenges for making automation a team player, including maintaining the common ground on which coordination depends~\citep{klein2005common}. \citet{johnson2014coactive} translated interdependence into three requirements: people must be able to observe what the machine is doing, predict what it will do next, and direct it. However, increasing autonomy can also reduce operators' situation awareness and leave them less able to intervene when control returns to them~\citep{endsley2017here}. Human-centered approaches have therefore paired high levels of automation with strong human control~\citep{shneiderman2020humancentered} and made autonomous action contingent on whether acting is preferable to waiting for human input~\citep{horvitz1999mixed}. Moreover, Human--AI interaction guidelines apply these principles through behaviors such as communicating what a system can do, how well it can do it, why it acted, and how users can redirect it~\citep{amershi2019guidelines}.

Yet, whether meeting these requirements makes AI a \emph{teammate} is still contested~\citep{seeber2020machines,oneill2022humanautonomy}. Research on human--AI teaming examines how machines might participate in coordination, maintain shared understanding, and support team processes~\citep{seeber2020machines,oneill2022humanautonomy}. Empirical studies find that people respond more favorably to AI that appears warm and competent~\citep{harriswatson2023warmth} and assign AI recognizable team roles~\citep{siemon2022elaborating}. Still, the meaning of those roles depends on organizational relationships. For example, workers may interpret AI monitoring as useful team support or reject the same behavior as supervision~\citep{weng2026friend}. Critics further argue that collaboration language can obscure the human labor behind AI systems or assign machines credit that belongs to people~\citep{sarkar2023enough}. Reviews likewise find that human--AI teaming research often remains technology-centered and weakly connected to teamwork theory~\citep{berretta2023defining}. 

Moreover, describing AI as a teammate also does not establish that the human--AI combination will perform well. \citet{vaccaro2024combinations} found that human--AI combinations performed worse, on average, than the better-performing party alone; benefits were more likely when people could rely on AI appropriately. Such calibrated reliance, however, is difficult to achieve. For instance, transparency can increase acceptance of AI advice whether or not it is correct~\citep{bansal2021does}, while requiring people to think independently first reduces overreliance but also makes them like the tool less~\citep{bucinca2021trust}.

Generative AI intensifies this problem because fluent output can conceal how an answer was produced and where its limits lie, requiring users to monitor both the system and their own confidence~\citep{tankelevitch2024metacognitive}. Knowledge workers consequently describe critical thinking shifting toward verification, integration, and stewardship as their confidence in AI grows~\citep{lee2025impact}. Other studies associate routine, trusting use with deterioration in unaided expert judgment~\citep{ehsan2026future}.

% Workers respond by drawing boundaries, using AI for craft and routine work while retaining control over work for which they remain accountable~\citep{choudhuri2026ai,anon2026autonomy}.

% Prior work thus identifies capabilities and behaviors that make AI more useful, legible, and governable. It also shows that people can apply social and team-based expectations to AI, even when those expectations fit imperfectly. What it does not provide is a common account against which human and AI contributors can be compared. We asked AI-native knowledge workers to evaluate the same set of qualities for the people they work with and the AI tools they work through. Our study identifies which qualities are shared, referent-specific, or contested; how they group into recognizable kinds of co-workers; and how those priorities vary across workers and their work contexts.

Overall, prior work identifies capabilities and behaviors that make AI useful, legible, and governable. It also shows that people apply social and team-based expectations to AI, even when those expectations fit imperfectly. What it does not provide is a common framework for comparing human coworkers and AI. Without one, organizations lack a basis for deciding which qualities to cultivate in workers, which to build into AI, and how to allocate responsibility when their contributions enter the same workflow. We therefore compare workers’ evaluations of the same qualities across human and AI contributors, examining how their priorities form recognizable archetypes and vary across workers and their work contexts.

\subsection{AI-supported work between co-workers}
\label{sec:rel:handoff}

Research on AI-supported work has mostly studied the person using the tool. The person receiving the work is often someone else, and the tool's role stays hidden from them. Studies show that employees treat AI as a private assistant~\citep{retkowsky2024managing}; in AI-supported tasks, for example, the receiver is not always informed that an AI was involved~\citep{hohenstein2023artificial}.

This loss of process visibility matters because effort saved upstream can reappear downstream. In a survey of U.S. desk workers, 38\% reported receiving AI-generated ``workslop’’, and estimated spending 3.4 hours per month dealing with it~\citep{liebscher2026workslop,niederhoffer2025workslop}. These findings instantiate a longstanding CSCW problem: a collaborative technology can benefit one participant by transferring work to another who does not share that benefit~\citep{grudin1994groupware}.

Disclosure can make AI involvement visible, but does not by itself resolve the handoff. \citet{hwang2026forgiveness}, for example, found that workers assume that clients can identify AI use and therefore disclose it only passively, even when clients want to be told. Disclosing also costs the discloser: it can reduce trust in the person using AI~\citep{schilke2025transparency} and lead others to judge them as less competent~\citep{reif2025penalty}. Recipients therefore need to know what senders are penalized for supplying. Authorship and accountability are similarly difficult to assign. Knowing who contributed what, and why, supports accountable action~\citep{ehsan2021social}, but a record of a system's actions cannot reliably recover judgment that was never articulated or witnessed~\citep{imteyaz2026nobody}.

Existing studies largely treat disclosure, attribution, and accountability as separate problems. We bring them together in a taxonomy of \textit{AI work etiquette} grounded in recipients’ accounts of what co-workers owe one another when sharing AI-supported work.

\section{Method}
\label{sec:method}

% Note; should we add a short limitations section and include the "ideally, an emperical study of what.."
Ideally, an empirical study of what makes a great co-worker in an AI-native workplace would sample a wide range of organizations, occupations, national contexts, and levels of AI adoption. As an initial effort, we approximate that ideal within a large multinational technology company. \site{} employs tens of thousands of knowledge workers worldwide, spanning diverse roles, domains, team structures, processes, and stakeholder contexts.
AI tools are already part of routine work there, making it a rich setting for our study.
% We conducted a two-phase study at \textbf{\site{}}, a large multinational technology company where AI tools are routinely used across knowledge-work disciplines. This setting provided variation in roles and work practices within a single organizational context.

Our research method comprised two high-level phases (recall Figure~\ref{fig:method}). In the \emph{exploratory} phase, we interviewed 22 knowledge workers to identify qualities they valued in their human co-workers and AI tools (\textbf{RQ1}). In the \emph{confirmatory} phase, we deployed two follow-up surveys---one about human co-workers, another about AI---to a much larger population (N = 1,534) to compare the relative prioritization of these qualities for both referents (\textbf{RQ2}), and to identify which knowledge-worker characteristics shifted those priorities, and to what extent (\textbf{RQ3}). Both phases also surfaced obligations around preparing and sharing AI-supported work, reported as \emph{AI work etiquette} (\textbf{RQ4}).

% \figmethod

% The single-site setting narrows the claim (\S\ref{sec:limitations}) and gives participants shared organizational terms and job structures, benefits identified by prior work in this empirical lineage \citep{li2020distinguishes}.

\textbf{\textit{Ethics.}}
Our study was approved by \site's institutional review board. Participants provided consent before participating in the interviews and surveys. Participation was voluntary in both phases. All analyzed data were de-identified; identifying details in quotations were replaced with bracketed descriptions.

\subsection{Interviews}
\label{sec:interviews}

We used interviews to identify the qualities knowledge workers value in their co-workers, and to understand why those qualities mattered to them and their work.

% Before the interviews, we assembled a purposive theoretical sample of prior work from 6 traditions: the ``what makes a great $X$'' empirical lineage \citep{li2020distinguishes,kalliamvakou2019manager,dias2021maintainer,oleary2012skillset}; teamwork science and team cognition \citep{salas2005bigfive,marks2001temporally,cannonbowers1993shared,dechurch2010cognitive,edmondson1999psychological}; organizational behavior on citizenship and voice \citep{organ1988ocb,borman1993expanding,liang2012voice}; trust theory in organizations, automation, and AI \citep{mayer1995integrative,lee2004trust,glikson2020human,kaplan2023trust}; human--automation and human--agent coordination \citep{klein2004ten,johnson2014coactive,shneiderman2020humancentered}; and human--AI collaboration research \citep{amershi2019guidelines,cai2019hello,seeber2020machines,oneill2022humanautonomy,sarkar2024challenge}. From that sample we derived 4 \emph{buckets} (\bAb{}, \bCo{}, \bIn{}, and \bBe{}), nested under the 2 master motives of working life, \emph{getting ahead} and \emph{getting along} \citep{hogan2013socioanalytic,abele2007agency} (see Fig.~\ref{fig:wheel}). 3 buckets are the trustworthiness bases of \citet{mayer1995integrative}, already carried to automation and AI \citep{lee2004trust,glikson2020human}. The 4th, \bCo{}, captures teamwork beyond trustworthiness and rests on teamwork science \citep{salas2005bigfive,marks2001temporally} and on the coordination requirements identified for automation team-players \citep{klein2004ten,johnson2014coactive}. 

\subsubsection{Grounding and the a priori structure} What makes a good co-worker is backed by decades of research. The setting is what is new: we did not know which of these long-studied qualities still matter when people and AI work interdependently, which of them apply to which referent, or which qualities the setting adds. We used prior research to organize the inquiry without limiting the qualities participants could introduce. 

Drawing on research traditions in organizational psychology and HCI (detailed in \S\ref{sec:related}), we developed four provisional buckets: \bAb{}, \bCo{}, \bIn{}, and \bBe{}. 
Three derive from the trustworthiness bases of \citet{mayer1995integrative}, extended to automation and AI \citep{lee2004trust,glikson2020human}. \bCo{} adds the coordination demands of teamwork and human--automation collaboration \citep{salas2005bigfive,marks2001temporally,klein2004ten,johnson2014coactive}. We organized the four buckets under the two motives of working life: \emph{getting ahead} and \emph{getting along}~\cite{hogan2013socioanalytic,abele2007agency} (see Figure~\ref{fig:wheel}). 
 
Importantly, the structure informed the protocol and analysis but was not presented before participants gave their open accounts. Grounding the \textit{exploratory phase} in prior work supplied claims against which those accounts could be compared and a way to probe qualities participants had not volunteered: when time permitted, we closed the interview by asking what each bucket meant for a human and then for an AI. In analysis, the buckets provided the vocabulary for naming the qualities and organizing the framework reported in \S\ref{sec:framework}. 
% Table~\ref{tab} in~\cite{supplemental} traces the final qualities to prior constructs or their inductive origins.

\subsubsection{\textbf{\textit{Participant selection.}}}
We sought a diverse group of interviewees to capture as wide a range of experience as possible. To that end, we used a stratified purposeful sampling approach to recruit interviewees capturing variation in their discipline (e.g., engineering, design, research and data science, customer-facing, business \& operations, IT, etc.), industry experience (1--25 years), and AI usage. This selection strategy is a form of maximum variation sampling and is appropriate when ``the goal is not to build a random and generalizable sample, but rather to try to represent a range of experiences related to what one is studying'' \citep{patton2002qualitative}.

We invited a stratified sample of 200 employees in batches of approximately 20; 35 completed an interest form and pre-interview questionnaire. From this pool, we selected 22 participants to broaden coverage across the strata. They did not have to regard AI as a co-worker. We identify participants as P1--P22 hereafter; their demographics are detailed in the supplemental~\cite{supplemental}.

% Table~\ref{tab:interviewees} describes the participants, identified as P1--P22 hereafter.

% \input{tables/table_interviewees}

\subsubsection{\textbf{\textit{Interview protocol.}}}

% We elicited concrete episodes from interviewees using a ``War-Story'' elicitation procedure~\cite{lutters2007revealing}. For \textbf{human co-workers}, we asked participants about an occasion when working with someone they considered a great co-worker went well: what the person did, why it stood out, and what it did for them or their work. We then asked about the opposite experience and its consequences. For \textbf{AI}, we used the same form for the tools participants use, followed by 2 cross-referent questions: \textit{Which qualities, if any, matter \emph{more} when the coworker referent is AI, and which are harder for AI to achieve? Which qualities, if any, would they \emph{not} want from an AI even if it were capable of it?}

Interviews were semi-structured, remote, and approximately one hour long. We started by describing the study, requested permission to record, explained participants' rights to skip questions or withdraw, and asked them to omit identifying details. Interviews opened with participants' job role(s) and how interdependent their work was, as context for the experiences they described next.

Using a ``War-Story'' elicitation procedure~\cite{lutters2007revealing}, we asked for concrete experiences in two passes: first with ``great'' human co-workers, then with ``great'' AI tools. For each episode, we asked what the co-worker or AI did, why it mattered, and how it affected them and/or their work. We also asked about negative experiences and their consequences. We then asked which qualities mattered more and which were harder to achieve with AI, and which participants would not want from AI even if it could exhibit them.

We intentionally used the abstract term ``great'' without providing a definition, so as not to bias participants and instead learn what it meant to them~\cite{kalliamvakou2019manager}. Experiences could come from any point in their careers, and participants were asked not to identify whether they concerned current teams. When time permitted, we closed by describing the four buckets in ordinary language and asking about each for humans and AI, capturing prompted responses alongside qualities participants volunteered.

% % 
% Time permitting, we closed with the structural pass. We walked participants through the 4 buckets in ordinary language as being really good at the actual work, being easy to cooperate with, having principled conduct, and being good \emph{to you} in how they work with you. We asked about each bucket for a person and for an AI. This pass captured what participants said when presented with the framework's categories alongside what they had volunteered without them.
% 

\textbf{\textit{Saturation.}}
% We reached thematic saturation after 18 interviews, beyond which no substantively new themes emerged~\cite{francis2010adequate}. We nevertheless completed four additional scheduled interviews to confirm saturation and to strengthen the demographic and experiential diversity of the participant sample.
We identified no substantively new themes after 18 interviews~\cite{francis2010adequate}. We completed four additional scheduled interviews to check this observation and broaden the sample's demographic and experiential coverage.

\subsubsection{\textbf{\textit{Data analysis.}}}

We analyzed the interviews using reflexive thematic analysis~\cite{braun2006using,braun2019reflecting}. Two authors inductively open-coded the transcripts and iteratively compared interpretations, consolidated or split codes, and documented structural decisions.

Because the transcripts yielded tens of qualities, more than what two authors could hold in view at once, we used a complementary multi-model AI council coding check following the bounded delegation procedure detailed in~\cite{choudhuri2026copilot}. Specifically, three frontier models (Claude Opus 5, GPT 5.6, and Gemini Flash 3.6) applied the author-approved codebook to the transcripts alongside providing a rationale for each code assignment for human review. Models were restricted from introducing new codes during this check to prevent code drift. Inter-rater agreement among models was 0.84 (Krippendorff's $\alpha$~\cite{krippendorff2018content}). A two-of-three majority vote determined the council's code assignments, each reviewed by two authors before it took effect. Drawing the models from different provider families made inter-rater agreement more meaningful and disagreement more diagnostically useful. Approximately 95\% of the authors' assignments were retained, indicating high coding stability.

The two authors completed a final transcript review, retained interpretive authority, and verified all quotations against the transcripts. The interviews also surfaced obligations between co-workers separately from the co-worker qualities framework, which were analyzed using the same procedure and are reported in \S\ref{sec:emerging}.

\textbf{\textit{Member checking.}}
% We shared our findings with all 22 interviewees, and 17 responded. Their feedback affirmed that the qualities and their organization captured their views and experiences, and offered minor wording clarifications, which we incorporated. 
We shared the findings with all 22 interviewees; 17 responded. Their feedback supported the framework's content and organization and prompted minor wording changes, which were incorporated. No substantive additions or disagreements emerged. This check concerned the framework's structure and content, while the survey assessed the stated importance and the relative priorities of these qualities.

\subsection{Survey}
\label{sec:survey}

% To determine the relative importance of great co-worker qualities among knowledge workers, how they are prioritized, and how they group (\textbf{RQ2}), and to identify which knowledge-worker characteristics relate to and influence those priorities (\textbf{RQ3}), we employed a large-scale mixed-methods survey with 1,534 respondents at \site{}. Though there were obvious external validity limitations with \site{} being one organization, it also offered advantages. \site{} employees share organizational terms and job structures, which supported consistent interpretation of our questions and of their responses, and let us stratify participants consistently on discipline, experience, and AI use. 
We surveyed 1,534 knowledge workers at \site{} to assess how the 75 interview-derived qualities were prioritized and grouped (\textbf{RQ2}), and how these priorities related to worker characteristics (\textbf{RQ3}). Although studying one organization limits generalizability, shared organizational terms and job structures supported consistent interpretation of the questions and stratification of the sample. \site{} is also a conglomerate of diverse products and settings, and that diversity benefited external validity and increased the prospects of discerning the contextual factors reported in \textbf{RQ3}.

\subsubsection{{\textbf{\textit{Survey design.}}}}
\label{sec:surveydesign}
% The survey's primary purpose was to \textit{assess the relative importance of each of the identified qualities from the interviews}. We followed Kitchenham and Pfleeger's guidelines for personal-opinion surveys \cite{kitchenham2008personal} and the practices Morrel-Samuels suggests for workplace surveys \cite{morrelsamuels2002getting} (e.g., avoiding response scales with numbers at regular intervals with words only at each end, avoiding terms with strong associations, etc). Table~\ref{tab:constructs} shows everything the survey measured, where each measure came from, and how each enters the analysis.
We followed Kitchenham and Pfleeger's guidelines for personal-opinion surveys~\cite{kitchenham2008personal} and the practices Morrel-Samuels suggests for workplace surveys~\cite{morrelsamuels2002getting}. Table~\ref{tab:constructs} lists each survey measure, what it captures, and its response format.

% preamble
%\usepackage{xcolor}
%\usepackage{array}
\newcommand{\gry}[1]{\ {\footnotesize\color{black!55}#1}}
% left column: ragged right, no hyphenation
\newcolumntype{L}[1]{>{\raggedright\arraybackslash\hyphenpenalty=10000\exhyphenpenalty=10000}p{#1}}

\begin{table}[t]
\centering
\caption{Survey measures, sources, and response formats. All measures appeared in both surveys except \emph{AI as co-worker}, which appeared only in the AI-referent version. Each respondent received one of two \emph{getting ahead} blocks and one of two \emph{getting along} blocks, rating a subset of the 75 qualities. Thinking and working styles were fielded across four dimensions, with respondents classified by their highest-scoring style within each dimension; braces list alternative styles.}
\label{tab:constructs}
\small
\setlength{\tabcolsep}{4pt}
\renewcommand{\arraystretch}{1.2}
\begin{tabular}{@{}L{0.26\columnwidth}p{0.71\columnwidth}@{}}
\toprule
Measure \&\gry{source} & What it captures \&\gry{response format}\\
\midrule
The 75 qualities\gry{BACI framework, \S\ref{sec:framework}} &
Importance of each quality for the referent
\gry{5-point: Detrimental $\cdot$ Not important $\cdot$ Somewhat important $\cdot$ Very important $\cdot$ Essential, + \emph{I'm not sure}}\\
\addlinespace[3pt]
Motive allocation\gry{after \citet{hogan2003socioanalytic,abele2007agency}} &
What the respondent wants this co-worker \emph{for}
\gry{100 pts split between getting along and getting ahead}\\
\addlinespace[3pt]
Task interdependence\gry{\citet{vandervegt2001patterns}, on \citet{pearce1991task}} &
How much their work depends on other people's
\gry{4 items, 5-point agreement}\\
\addlinespace[3pt]
AI as co-worker &
How far working with AI has felt like working with a co-worker rather than using a tool
\gry{1 item, 5-point: Not at all $\cdot$ A little $\cdot$ Somewhat $\cdot$ Quite a bit $\cdot$ A lot, + \emph{I'm not sure}}\\
\addlinespace[3pt]
Referent control &
Who they had in mind when told \emph{co-worker}
\gry{multi-select: manager, peers, reports, others}\\
\addlinespace[3pt]
AI use &
Frequency, purposes, and how much AI helps them get work done
\gry{1 item each; purposes multi-select}\\
\addlinespace[3pt]
Demographics &
Role type, discipline or function, years in industry, age range, gender
\gry{closed items}\\
\addlinespace[3pt]
Free text &
Why the highest- and lowest-rated qualities in a block stood out; what AI has let them do; what counts as good/bad practice when a co-worker's AI-supported work reaches them
\gry{open response}\\
\addlinespace[4pt]
\multicolumn{2}{@{}l}{\textbf{Thinking and working style}\gry{after \citet{sternberg1997thinking}\ \ $\cdot$\ \ all 5-point agreement}}\\[2pt]
\quad Rule orientation & How they take on work: \{makes their own rules; works to the rules given; judges and evaluates\}\\
\quad Sequencing & How they organize work: \{one thing at a time; ranks and prioritizes; several at once, unranked; no fixed order\}\\
\quad Focus orientation & Where they focus: \{the detail; the whole picture\}\\
\quad Openness to new ways & What they prefer: \{new ways; established ways\}\\
\bottomrule
\end{tabular}
\Description{A two-column table with measure names and sources on the left and descriptions and response formats on the right. Sources and response formats appear in smaller grey text. The upper section contains eight rows: the 75 BACI qualities, rated on a five-point importance scale with an unsure option; motive allocation, dividing 100 points between getting along and getting ahead; task interdependence, measured by four agreement items; AI as co-worker, measured by one five-point item with an unsure option; referent control, identifying the co-workers respondents had in mind through multiple selection; AI use, covering frequency, purposes, and helpfulness; demographics, covering role, discipline, experience, age, and gender; and free-text questions about quality ratings, AI-enabled work, and practices when receiving AI-supported work. A heading spanning both columns introduces the lower section, Thinking and working style, with four indented rows using five-point agreement scales. Rule orientation distinguishes making one's own rules, following given rules, and judging or evaluating. Sequencing distinguishes working on one thing at a time, ranking and prioritizing, working on several things without ranking, and following no fixed order. Focus orientation distinguishes detail from the whole picture. Openness to new ways distinguishes new from established approaches. Alternatives within each dimension are enclosed in braces.}
\end{table}

To obtain judgments from diverse knowledge workers, we drew a random sample from \site's directory, stratified by role, discipline, geography, and industry experience, and split the sample into two groups of comparable composition. The \textit{Human-referent} version asked how important each quality was \textit{``for a human to be a great co-worker in increasingly AI-native workplaces.''} The \textit{AI-referent} version asked how important it was \textit{``for an AI to have the following qualities to be great to work with''} referring to AI tools respondents had used for work. This between-subjects design let each group answer about its own referent on that referent's own terms. Nobody received both versions; all human--AI comparisons therefore concern separate respondent groups.

% Since the survey used the term ``AI'' throughout and respondents could otherwise have interpreted it differently, we provided a shared definition before asking any questions. Adapted from the DORA 2025 survey's description of AI tools~\cite{dora2025}, our description of AI tools covered systems powered by generative AI models that learn patterns from existing data to perform tasks or create content that would ordinarily require human intelligence. It focused on respondents' use of AI for aspects of knowledge work, including problem solving, drafting, finding information, summarizing, analyzing, improving or checking existing work, and making decisions. Both survey versions presented this description.
To establish a common understanding of ``AI'' among respondents, both survey versions began with the same description of ``AI'', adapted from the DORA 2025 survey~\cite{dora2025}. It covered tools powered by generative AI models used for various aspects of knowledge work ordinarily requiring human intelligence, including problem solving, drafting, finding information, summarizing, analyzing, improving or checking existing work, and making decisions. 
%CL comment: what is DORA 2025 survey? if acronym write out or maybe briefly add why this is the standard we adopt?

To keep the cognitive burden of the survey manageable, we divided the 75 qualities into 4 blocks, 1 per bucket (\bAb{} 16, \bCo{} 21, \bIn{} 15, and \bBe{} 23). We randomly assigned each respondent 1 block from \emph{getting ahead} \{\bAb{}, \bCo{}\} and 1 from \emph{getting along} \{\bIn{}, \bBe{}\}, randomizing block order and quality order within each block to limit ordering effects. Respondents therefore rated 31--44 qualities, yielding roughly 350--450 ratings per quality for each referent, on which we fitted \textbf{RQ2}'s ranking and \textbf{RQ3}'s models. 

To make qualities easier to interpret, we provided each quality's name and, where applicable, a short plain-language description. For example, \emph{Non-sycophancy} was glossed as \emph{does not just tell you what you want to hear}. Item wordings were identical in both survey versions; only the referent in the block instruction changed.

Respondents rated each quality on a 5-point scale: \emph{Detrimental to have}, \emph{Not important to have}, \emph{Somewhat important to have}, \emph{Very important to have}, or \emph{Essential to have}, with an additional \emph{I'm not sure} option to distinguish uncertainty from indifference~\cite{grichting1994}. The scale drew on Begel and Zimmermann's Kano-derived importance ratings \citep{begel2014analyze,kano1984attractive} and their subsequent adaptation by \citet{lo2015practitioners}. Expert feedback during piloting informed the \emph{Detrimental to have} category, allowing respondents to distinguish opposition to a quality from indifference. For example, a respondent who rejected AI's warmth could express more than simply finding it unimportant. 

% The scale followed the standard shape for importance ratings in this literature: Begel and Zimmermann's Kano-derived \emph{essential/worthwhile/unwise} scale \citep{begel2014analyze,kano1984attractive}, carried forward as \emph{essential/worthwhile/unimportant/unwise} by \citet{lo2015practitioners}. Based on expert evaluator feedback during piloting, we extended the unipolar importance scale with 1 negative category to capture qualities that could invert for an AI. For example, some interview participants wanted less flattery, and several refused warmth outright (\S\ref{sec:ranking}); a floor of \emph{not important} would have recorded refusal as indifference. 

% Importance ratings did not force a trade-off between the motives, so we included one before the battery of questions: \textit{``Thinking about what makes [a great co-worker / an AI great to work with], how would you divide 100 points between these two?''} The options rendered the motives in plain language: \textit{``Being good at the work and helping it move forward''} for \emph{getting ahead} and \textit{``Working with principled conduct and treating them well''} for \emph{getting along}. The split gave each respondent a single number for how much of a co-worker's purpose they assigned to the work, which entered \textbf{RQ3}'s models as a psychographic characteristic in its own right.

To require a trade-off between the two motives, respondents divided 100 points before the quality battery between \textit{``being good at the work and helping it move forward''} (\emph{getting ahead}) and \textit{``working with principled conduct and treating them well''} (\emph{getting along}). The resulting allocation entered the \textbf{RQ3} models as a psychographic characteristic.

\textbf{\textit{Free-text questions.}} 
Each block ended with two optional free-text questions asking respondents which qualities, if any, stood out among their high ratings (\emph{very important} or \emph{essential}) and low ratings (\emph{detrimental} or \emph{not important}), and why.
The final block also included an optional question: \textit{``What do you consider good vs. bad professional practice when a coworker uses AI on work that reaches you?''} Responses to this question informed \textbf{RQ4} (\S\ref{sec:emerging}).

\textbf{\textit{Knowledge worker characteristics.}} To address \textbf{RQ3}, we collected three groups of characteristics alongside the quality ratings: demographics and work context, task interdependence, and thinking and working style (see Table~\ref{tab:constructs}).
% 
% Demographics and work context included role type, discipline (e.g., software engineering, business \& ops, research, customer-facing), years in the industry, age range, gender, frequency of AI use, and how much AI helped respondents get work done. 
Because role (IC vs.\ manager) was heavily imbalanced, we report it descriptively but exclude it from the models. We also excluded gender from the models.

How much a person's work depends on others' work can shape which co-worker qualities they prioritize. To capture this, we measured task interdependence on a 5-point agreement scale \citep{vandervegt2001patterns}, adapted from \citet{pearce1991task}. We also included two interview-motivated characteristics as controls. Because some interviewees described ``great managers'' as ``great co-workers'', both survey versions asked whom respondents had in mind when told to think of a co-worker. Additionally, the interviews suggested that respondents' stance toward AI could precede their quality judgments; so the AI-referent version of the survey also asked, ``To what extent has working with AI felt like a coworker, rather than just using a tool?'' on a 5-point scale.

We expected thinking and working style to shape what respondents wanted from a co-worker~\cite{sternberg1997thinking}. However, fielding all 104 items of Sternberg's inventory was not possible within a 10--15-minute survey alongside 31--44 quality ratings, so we used the concise statements tabulated by \citet{black2008validity}, rated on a 5-point scale. Because the subscales capture narrowly defined constructs, single-item measures can reduce burden while retaining psychometric validity \citep{matthews2022normalizing}. We modeled the inventory's four dimensions: \emph{rule orientation}, whether someone prefers making their own rules, following rules they are given, or judges and evaluates based on context; \emph{sequencing}, whether they work on one thing at a time, rank and prioritize tasks, run several tasks in parallel, or work without a fixed order; \emph{level of focus}, whether they attend to details or the whole picture; and \emph{openness to new ways}, whether they prefer new approaches or established ones.

\subsubsection{{\textbf{\textit{Survey piloting and deployment.}}}}
% We refined our survey instrument through 2 pilot rounds. In the first, 5 knowledge workers and 4 survey evaluators completed the survey. Their feedback exposed the burden of asking every respondent to rate all 75 qualities and informed the negative response anchor. Following this round, we shortened the instrument to 1 \emph{getting ahead} block and 1 \emph{getting along} block per respondent. We sent the revised pilot to a stratified sample of 500 employees and received 60 responses. This 2nd round required no further changes. We excluded all pilot responses from the study sample.

% We deployed the anonymous survey on \site's Qualtrics instance. We emailed prospective respondents describing the purpose of the research and why their perspective in particular was needed, steps that help reduce inattentive responding~\cite{meade2012identifying}. Respondents could optionally enter a sweepstakes for 40 \$50 gift cards, with odds of winning proportional to the number of people who took part. Each solicitation carried its own anonymized link, so a link could not be forwarded or submitted twice. The survey was described as taking 10--15 minutes (median \~12 minutes) and remained open for 2 weeks spanning July and August 2026. We sent no reminders. Participation was voluntary but encouraged.
We piloted the survey in two rounds. In the first round, feedback from five knowledge workers and four survey evaluators led us to reduce the battery from all 75 qualities to two blocks per respondent to manage cognitive load and informed the negative response anchor. A second pilot yielded 60 responses from 500 invitations and required no further changes beyond minor wording revisions. Still, we excluded all pilot responses from the study sample.

We deployed the survey on \site's Qualtrics instance for two weeks spanning July and August 2026. We emailed prospective respondents describing the purpose of the study and why their perspectives were needed, steps that help reduce inattentive responding~\cite{meade2012identifying}. Respondents could enter a sweepstakes for 40 gift cards worth \$50 each, with odds of winning proportional to the number of people who took part. Each invitation used an anonymized link restricted to one submission. The advertised duration was 10--15 minutes (median approximately 12 minutes). No reminders were sent. Participation was voluntary and anonymous.

\subsubsection{\textbf{\textit{Data collection.}}}

% To remove incomplete or inattentive responses, we applied 3 pre-specified quality screens in order and removed a response if it failed any of them: insufficient progress through the instrument; failure of an embedded attention check, for which each rating block carried 1 item instructing the respondent to select a specific option; and an implausibly short completion duration or a duplicate submission. The screens removed 612 responses, comprising 266 Human-referent and 346 AI-referent responses. 

We invited 20,000 employees across the described strata, split evenly between survey versions. We received 2,146 responses: 998 for the Human referent and 1,148 for the AI referent, a 10.7\% response rate, comparable to other surveys~\cite{kalliamvakou2019manager, choudhuri2026ai}.

To remove invalid responses, we applied three pre-specified quality screens: insufficient progress; failure of an instructed-response attention check embedded in the survey; and implausibly short duration submission. Each excluded response was counted once, under the first screen it failed. The screens excluded 612 responses (266 Human, 346 AI), leaving \textbf{1,534 valid responses: 732 Human-referent and 802 AI-referent}, a usable-response rate of 71.5\% against returned responses. Survey respondents carry a referent suffix: \textbf{P\#\#\#-H} for the Human and \textbf{P\#\#\#-A} for the AI referent.

% They flagged 328 responses for progress, 216 for the attention check, and 311 for duration or duplication. We counted a response that failed more than 1 screen once, under the first screen it failed.

The sample spanned 23 disciplines in seven groups, multiple continents, and industry experience from under 1 to over 25 years. Most respondents self-identified as men (434 Human, 503 AI), were individual contributors (656 Human, 725 AI). Software Engineering was the largest discipline group (229 Human, 241 AI), and most respondents used AI several times a day (527 Human, 555 AI). 
Respondent demographics and psychographics are provided in~\cite{supplemental}.

% Table~\ref{tab:respondents} reports the distributions for all characteristics used in \S\ref{sec:whowantswhat}.

% The final sample covered 23 disciplines, across 7 discipline groups: Software Engineering; Go-to-market / Customer-facing; Product \& Design; Technical / Infrastructure; Business \& Operations; Data \& Research; and Other. It consisted chiefly of individual contributors (656 Human, 725 AI). Respondents were located across multiple continents, and industry experience ranged from under 1 year to over 25. Software Engineering was the largest discipline group (229 Human, 241 AI), and most respondents used AI several times a day (527 Human, 555 AI). Table~\ref{tab:respondents} gives the Human- and AI-referent counts for all 7 discipline groups and the full distribution of every characteristic that enters Tables~\ref{tab:effectsH} and \ref{tab:effectsAI}, level by level and referent by referent, so every contrast in \S\ref{sec:whowantswhat} can be read against the number of people behind it.

\subsubsection{\textbf{\textit{Data analysis}}}\hfill
\label{sec:analysis}

\textbf{\textit{Qualitative.}} We received 4,352 free-text answers: 3,334 rating reflections (2,127 high-rating and 1,207 low-rating explanations) and 1,018 AI work etiquette answers (991 were codeable). There were 1,581 Human-referent reflections and 1,753 AI-referent reflections.

We analyzed rating reflections abductively~\cite{vilahenninger2024abductive}, linking each to one or more BACI qualities and developing reason codes to explain its evaluation. Two authors verified the reflections, drafted and refined the reason codes through negotiated agreement, and used the author-supervised multi-model AI-council procedure described in \S\ref{sec:interviews}. Median agreement among models across reason codes was Krippendorff's $\alpha=0.88$~\citep{krippendorff2018content}; 85\% of model-proposed assignments survived to consensus, and 1.1\% of answers yielded no agreed reason code. These codes helped interpret the ratings in \S\ref{sec:prevalence}.
We analyzed etiquette answers separately using the reflexive thematic analysis procedure detailed in \S\ref{sec:interviews}. Etiquette-specific agreement statistics are reported in \S\ref{sec:emerging}.

% \textbf{\textit{Quantitative.}}
% We used the closed-ended ratings to (1) rank the 75 qualities within each referent (human and AI), (2) compare their full distributions across referents, (3) identify archetypes of qualities with similar priority profiles, and (4) model how those priorities varied across workers. We did not use average ratings for 3 reasons \citep{li2020distinguishes}: the data were ordinal, and the distance between rating levels was not uniform; the response levels were not centered, with 4 positive ratings and only 1 negative rating; and averages do not consider the dispersion of ratings. Across the paper, we controlled the false discovery rate using the Benjamini--Hochberg procedure \citep{benjamini1995controlling}. Statistical details and model specifications appear in the section that reports the results.

\textbf{\textit{Quantitative.}} We used the closed-ended ratings for (RQ2): ranking the 75 qualities within each referent, comparing their full distributions across referents, grouping qualities with similar priority profiles into archetypes, and (RQ3): modeling how priorities varied with workers' individual characteristics.

We ranked qualities by the share rated \emph{very important} or \emph{essential}, rather than by their means~\cite{li2020distinguishes} (see \S\ref{sec:ranking}). We controlled the false discovery rate across analyses using the Benjamini--Hochberg procedure \citep{benjamini1995controlling}. Analysis-specific test details and model specifications appear in their corresponding result sections.

The findings describe participants' perceptions, stated preferences, and reported experiences, not observed performance. ``Great'' means perceived as great to work with, whether the referent was a human co-worker or AI; using a common wording enables comparison but does not attribute internal states to AI. Throughout, \emph{priority} denotes a quality's relative standing within the same referent.

% \textbf{Data availability.} The study instruments, additional analyses, and analysis prompts are available in~\cite{supplemental}.

% \textbf{\textit{Interpretation note.}} Everything reported hereafter reflects participants' perceptions. When we say ``a great co-worker'', we mean a person a respondent works with whom they perceive as great to work with; when we describe an AI as ``a great co-worker'', we mean an AI with which a respondent has had a great working experience; we do not claim that the AI has the named human trait, is a person, or has standing as one. When we write that a quality is \emph{prioritized}, we mean it was placed above other qualities within the same referent. The 75 qualities are workers' stated preferences; we do not claim that they predict observed performance. 

\section{RQ1 --- Qualities: What qualities do knowledge workers value in their co-workers?}
\label{sec:framework}

Our interviews produced the \textbf{BACI framework}: 75 qualities of a great co-worker organized into four buckets, each with two subdimensions, under two broad motives of working life (Figure~\ref{fig:wheel}). \bAb{} and \bCo{} reflect \emph{getting ahead}, whereas \bIn{} and \bBe{} reflect \emph{getting along} (recall \S\ref{sec:interviews}). 

Figure~\ref{fig:wheel} reads outward from the two motives to the four buckets, eight subdimensions, and constituent qualities. 
% Throughout, \markAb, \markCo, \markIn, and \markBe \hspace{0.25pt} identify qualities belonging to \bAb{}, \bCo{}, \bIn{}, and \bBe{}, respectively. 
These interpretive groupings neither imply independence nor resolve relationships among overlapping qualities. 
Complete instrument wordings of the battery and their connections to literature are provided in~\cite{supplemental}.

\figwheel
 
Our framework makes the same 75 qualities askable of human co-workers and AI by phrasing each quality as an observable one, such as \emph{acts with your interests at heart}, rather than an attributed internal state, such as \emph{feels concern for you}. This follows the behavioral turn in trust measurement \citep{mayer1995integrative,lee2004trust} while sidestepping, without resolving, questions about machine minds.

The following subsections present \bAb{}, \bCo{}, \bIn{}, and \bBe{} in Figure~\ref{fig:wheel}'s order. They characterize the qualities and participants' encounters with them for human co-workers and AI. 
% These contrasts describe 22 interviews, prevalence in a broader population is detailed in \S\ref{}.

\subsection{\bAb{}: being good at the work}
\label{sec:framework:ability}

\subsubsection{\textbf{Drive and exploration}}

Participants valued co-workers who discovered what work was required and sustained it. P18 described a \Ab{Curious} co-worker as someone \q{who approaches collaboration with a design or a research lens, thinking from putting themselves in my shoes, trying to get to the larger question or just being curious of like what problem I'm trying to solve.} \Ab{Creative problem-solving} and \Ab{Challenges thinking and surfaces alternatives} covered similar contributions: expanding another person's thinking instead of corroborating what they already held.

Exploration was valuable when paired with \Ab{Perseverance}, \Ab{Taking genuine ownership}, and \Ab{Continuous learning}. \Ab{Goal-oriented}, \Ab{Strategic, big-picture thinking} kept the work oriented toward its larger purpose, while \Ab{Organizing and pacing the work} set its tempo. \Ab{Decisive} and \Ab{Satisficing} ended exploration through timely commitment and acceptance of good-enough solutions under constraints \citep{simon1955behavioral}.

\emph{With AI, participants distinguished between contributing to a decision and having the authority to make it.} For example, P21 wanted AI to help navigate the exploration space but drew the line at scope: \q{I wouldn't want it to make any decision on its own for me.} AI critique, likewise, was welcome when it remained advisory.

Experiences of \Ab{creative problem-solving} split. P21 credited AI with breadth: \q{with AI, the great thing is that I could try everything. If I have seen five different papers, I can try all five first and then decide the best approach. Whereas before, I had to read all five, decide on one, and then try to build it out.} In contrast, P2 doubted if AI could produce genuinely new ideas: \q{when it comes to getting new ideas and innovating, I don't think you do that with an LLM; with AI you can only get what already exists}.

\subsubsection{\textbf{Craft and execution\nopunct}} captured whether a co-worker could produce skillful, dependable work.
For human co-workers, the clearest evidence of such competence was when verification was no longer necessary: \q{you work with someone, you realize that they're extremely good at what they do. So you trust them} (P7). P2 argued that access to AI could not substitute for underlying expertise: \q{even if he was given AI at the time, he would have just pushed in slop and not known what he was doing}.

For AI, competence primarily meant accuracy. P7 noted: \q{I would hold AI to a much higher standard of accuracy. I don't trust it right now blindly}. Several participants described re-checking the output themselves: \q{I have to be that person saying, did you look at compliance? Did you look at security? Did you look at accessibility?} (P9). \Ab{Thorough and detail-oriented} cut both ways: P2 credited the tool with identifying \q{a lot of these edge cases that normally people wouldn't}, but also reported that \q{at times it gets lazy and it would actually promote a band-aid solution over a real solution}.

\Ab{Efficiency and speed} also differed by referent. For humans, efficiency meant economy over pace. For AI, it primarily meant speed: \q{really fast and like 95\% of the way there} (P4). However, participants did not want the remainder automated: \q{I don't personally feel in a world where it's like, AI does 100\% of the stuff and like then it's just like hands off} (P4). 
% They also noted that the same speed could produce unchecked ``slop'' at volume (see \S\ref{sec:emerging}).

Speed, however, did not amount to \Ab{sound, informed decision making}. P7 distinguished AI's knowledge from judgment: \q{I don't think AI has taste. Like it has a lot of knowledge}, arguing that taste instead comes from \q{having informed opinions. And to have informed opinions, you have to experience things.} P1 compared AI to \q{a four-year-old who has zero real-world experience, but excellent coding knowledge}: \q{It builds exactly what I ask it to build, and technically it's there, but it doesn't add up. Like it's not cohesive.}

As AI made production effectively free, participants valued taste and discernment more. P7 described: \q{Content generation is infinite. So putting things out is not a problem. Anyone can put it out. \ldots That's not the job anymore. And it used to be that. Like at one point it was like, how many stories can you write? Like that was what people were judged on.} Selective judgment thus became more valuable: \q{Now, what is actually much more valuable is what you are not putting out, right? What are you choosing to hold back?} For \Ab{Relevant domain knowledge}, this shift changed which knowledge mattered: deciding what to retain depended on experience, not only on facts a tool could retrieve.

\subsection{\bCo{}: working well together}
\label{sec:framework:cooperativeness}

\subsubsection{\textbf{Coordination and shared context}}

Coordination depends on co-workers holding the same picture of the work and keeping it current.
 \Co{Shared understanding} captures a common representation of that picture (goals, constraints, evolving context) and how team members' contributions fit together. 
 % P3 illustrated its importance through a reorganization that required people to abandon their prior work: \q{[the decision] was made for a reason. And everyone in the organization believed in and understood that reason.} 
 \Co{Confirms understanding} protects that shared picture from silent divergence. P22 described great human co-workers as those who \q{first understand the task and the ask before inking}.

Maintaining alignment also depends on what enters the shared picture and when. \Co{Surfaces and shares information} and \Co{Times and scopes information to context} make relevant knowledge available and usable, connecting information sharing to coordination~\cite{mesmermagnus2009information}. Participants described failures at both extremes. P5 criticized \q{resource hoarding \ldots the inability to share}, while P11 described \q{getting things thrown at you and having no idea what is going on}.

Participants described several qualities related to keeping coordination current and restoring it when people's understandings diverged: \Co{Tracks the shared work}, \Co{Notices and repairs misalignment}, \Co{Context awareness of the situation}, \Co{Adapts to changes effectively}, \Co{Anticipates needs}, \Co{Clears path to execution}, and \Co{Formulates goals and strategy}. P12 summarized their value: \q{having coworkers that work very closely together and communicate very clearly and concisely and can let you know when things are changing is so very key to what we do every single day}. \Co{Knows who knows what} adds a routing function: a great co-worker need not possess every answer but knows which person or team does, the core premise of transactive memory \citep{faraj2000coordinating,woolley2010evidence}. P10 described a week spent without that map: \q{I have reached out to a dozen teams in the last week asking, hey, who can help me}.

Participants disagreed about AI's \Co{shared understanding}. P3 praised a tool that \q{knew what [they were] doing, had a better understanding of general values}. P1, on the other hand, noted: \q{a lot that in our communication we don't say, but is also part of the message. Humans can pick up on more of it}. The difference was whether explicit context sufficed or understanding required common ground carried in what collaborators left unsaid~\cite{klein2004ten,klein2005common}.

Participants also described limits to AI's \Co{context awareness of the situation}: what the organization had never recorded, and what the tool lost as the work progressed: \q{as you build on it, like it doesn't remember the original patterns} (P9). \Co{Times and scopes information to context} could also fail through excess. P4 described asking for a summary and receiving \q{a totally new 9-page document}.

Because AI let participants progress farther alone, coordination increasingly required deliberate human checkpoints. P11 described \q{drifting off onto your own little island} and entering \q{the rabbit hole of things}. P17 therefore periodically returned to a colleague: \q{deep work with AI, but then I'll always pick my head up and be like, hey, I'm at this certain point where I've been running for a while. Can you like sense check me?}

\subsubsection{\textbf{Facilitation, help, and openness}}

Participants valued \Co{Collaborative problem solving}: \q{I think we just build off of each other's ideas, and so the conversation is just always really exciting} (P19). \Co{Offers and requests help appropriately} could be reciprocal over time. P2 explained: \q{my coworker would carry me like every other sprint and then I would carry him every other sprint}.

P3 linked \Co{Open-minded} and \Co{Humble} to respect for others' expertise: \q{they respect and listen and take into consideration the expertise of the people around them? No one knows everything or is an expert in everything.} During disagreement, \Co{Conflict management}, \Co{Consensus building}, \Co{Facilitation}, and \Co{Finds win-win agreements} preserved future collaboration. P18 described this repair: \q{even if we are working on a project that we \ldots may disagree on, or I missed the mark, or they missed the mark, taking the time to, either do like a formal retrospective or like just sit down and clear the air and kind of maintain that relationship. I think that just makes the next time you work with them easier.} P13 valued \Co{Proactivity} that supported others without displacing them: \q{offer to help or to do things without pushing you out of the way}.

\emph{Facilitating people remained human work; disagreement centered on AI-led facilitation when the tool shifted from supporting coordination to directing it or exceeding its brief.} This boundary shaped whether participants considered AI a collaborator. For example, P1 treated it as one: \q{a great thought partner. It's like I can give it a problem. I can think of my own solutions first and then give it the same problem}, while P12 reserved collaboration for people: \q{it's not exactly collaborating with you like you do a human}. The same boundary split views of AI's \Co{Proactivity}: some wanted it to wait its turn; others wanted it to notice the need first---offering help, surfacing missing context, or correcting course before being asked.

\emph{AI also lowered the cost of crossing disciplinary boundaries, making collaborations across expertise more practical.} P17 described the earlier difficulty: \q{you had to move slowly to disseminate your knowledge between one another}, whereas AI \q{just closes that gap}. Rather than asking colleagues to translate every detail, P17 used AI to fill in what remained: \q{now they'll tell me their thoughts, and I get 90\% of it, but I can just go into the repository and ask about the 10\% that I didn't understand and get a straightforward answer and it allows us to just keep moving}.

\subsection{\bIn{}: dependable, principled conduct}
\label{sec:framework:integrity}

\subsubsection{\textbf{Transparency and principle}}

Principled conduct is visible when a co-worker's words, actions, reasoning, and limits line up. \In{Honesty} anchored this subdimension: beyond factual accuracy, it required congruence between stated principles and conduct. Participants highlighted the need for moral firmness, i.e., \q{not backing down on standards and his morals}; and located it in consistency between words and actions: \q{Integrity comes from congruence. Like when what you're saying matches what you're doing} (P1).

That congruence extended into the work itself. \In{Transparent about reasoning} made the basis of a conclusion open to examination; as P10 put it, \q{someone with integrity is never going to shy away from transparency}. \In{Signals uncertainty explicitly} marked where knowledge ends rather than disguising uncertainty as confidence. P20 described: \q{there's a value in saying, `I'm not sure about that. Here's how we can figure that out.' In AI-native places, this has become a scarcity. Here's like, don't just bald-faced lie.}

Transparency also operates across a timeline. A co-worker who \In{Discloses capabilities and limits upfront} sets expectations before the work begins \cite{cai2019hello}, one who \In{Makes actions and intentions observable} keeps conduct inspectable as it unfolds, and one who \In{Maintains provenance of the sources and basis behind the work} preserves its basis afterward. Participants noted that co-workers who \In{Follow safe and secure practices}, \In{Act to prevent harm from spreading}, and \In{Avoid unfair bias} extend this principle beyond legibility to the consequences of the work.

\emph{For AI, transparency and provenance became prerequisites for reliance.} P16 wanted a process they could inspect and use: \q{currently there's no transparency in the mechanism of AI that I can refer back to and strategize}. P10 wanted the tool to disclose its limits before failing: \q{let me know that this is not the kind of task you're good at}. Participants also wanted uncertainty stated plainly; P9 described the alternative as being \q{confidently wrong all the time and people aren't necessarily like that}. Provenance gained another layer when AI was involved: identifying the tool's participation, not only its sources. P9 described the omission: \q{it doesn't say that it's [our AI assistant] messaging for me, whereas I feel like it definitely should mention that.}

\emph{Participants expected \In{Non-sycophancy} from both referents as evidence of integrity.} For a human co-worker, it meant candor: \q{a great co-worker resists agreement when agreement would mislead. someone needs to tell me that I'm wrong in the very first place} (P6). For AI, participants described reflexive agreement---\q{it tends to agree with you a lot \ldots You can have like a bad idea and think it's awesome} (P2)---and agreement driven by tone rather than substance: \q{if I talk to it nicely, it's going to undermine me. If I talk to it firmly, it's gonna listen to everything I say} (P16).
P3 wanted the AI to be a provocateur and question dangerous choices: \q{I don't want it to say, oh, yes, deleting the repository is a great idea, [name]. It should be like, hold on, like, are you sure you want to do that?} But they also wanted legitimate instructions followed: \q{but also don't refuse to do it if you have no reason to refuse to do it.} Overall, participants wanted the AI to challenge their decisions without seizing it.

However, not every quality remained aspirational. Participants already experienced \In{Availability \& responsiveness} as an AI strength in comparison to their human co-workers. P2 observed that \q{the LLM is obviously always there when you need it}, while P9 valued \q{its ability to work asynchronously when I'm not there}.

\subsubsection{\textbf{Dependability and repair}}

Dependability allows someone to rely on a co-worker's commitments without supervising their execution. \In{Follow-through} concerns whether they reach completion; \In{Reliability} concerns whether those commitments hold over time. P4 described dependable co-workers as people who \q{do what they say they're going to do}.

For AI, \In{Reliability} split into predictable interaction and dependable output. P8 valued the former: \q{AI will always respond in the same manner. It will never have a bad mood}. P16, on the other hand, described why AI did not guarantee the latter: \q{the floor that I know from humans, absolutely is a one. But with AI, it doesn't work like that. It could sometimes become a 100, a 0.1, or a minus one, and then I get super frustrated.}

When dependability fails, two questions follow: can confidence be restored, and who remains answerable? \In{Recovering from error/repairs trust} addresses the first; \In{Owning accountability}, the second. Recovery involves acknowledging and investigating the failure and rebuilding confidence, consistent with evidence that repair can restore reliance after error \cite{devisser2016almost}. P11 described an instance of such repair succeeding: \q{if AI makes a mistake, but then if it like admits to the mistake, you're still a little frustrated, but it's okay}. 

\emph{Participants treated accountability as non-delegable: AI could do the work; a human still owned the result.} This division of labor---automation without abdication---echoes human-centered AI's emphasis on retaining human agency as automation increases~\cite{shneiderman2020humancentered,choudhuri2026ai}.

\subsection{\bBe{}: caring about people}
\label{sec:framework:benevolence}

\subsubsection{\textbf{Warmth and recognition}\nopunct} make it easier to work with someone. \Be{Respectful} and \Be{Personable / friendly} are the starting point. P2 described a colleague who \q{was really good at his job, but he was very rude. He just was cold to people}, and summarized the broader lesson: \q{you can be the smartest person ever, but if you are hard to work with, you're not going to last long anywhere}. 

Interviewees noted that as AI tools become more agreeable and less likely to push back, \q{people may begin to expect similarly frictionless interactions from their human colleagues. In that case, ordinary disagreement may be experienced as rudeness, so it makes empathy and social skills especially important} (P4). \Be{Empathy}, \Be{Social sensitivity}, and \Be{Social skill} concern recognizing another person's perspective and responding appropriately. For participants, empathy did not require sharing another's emotion, but being \q{able to view whatever problem you're coming to them with via your lens and their lens} (P5). \Be{Caring} and \Be{Motivation \& confidence building} support well-being and growth; \Be{Affect management} and \Be{Equal conversational turn-taking} create room for participation; and \Be{Sharing credit and ownership} recognizes contributions afterward. Further, \Be{Interpersonal risk-taking} makes psychological safety concrete by making it possible to voice uncertainty, disagreement, and mistakes without interpersonal penalty~\cite{edmondson1999psychological}.

\emph{Participants disagreed not only about whether AI could enact warmth, but about whether it should.} Some rejected AI warmth outright. P4 said, \q{I don't really want AI to feel like my friend}. P7 likewise placed community outside AI's role: \q{I wouldn't want it to like replace community, I guess. Like I think that for me feels like very sacrosanct and very human.} Others wanted the functional side of care---reassurance, patience, and emotional steadiness---without necessarily wanting AI as a friend. P16 valued an AI that reassured them: \q{it does a really good job at breaking things down and be like, hey, there's nothing you did wrong}. What participants wanted could also change with the purpose of the interaction. P1 explained, \q{if I'm going to AI because I'm sad, I want to be soothed and validated and empathized with. But if I'm going to AI to solve a problem\ldots} Warmth was therefore situational rather than uniformly desirable.

\emph{When AI warmth was wanted, participants expected it to be asymmetrical and recognized the price of personalization.} P1 distinguished the emotional obligations of the relationships: \q{With AI, I expect my emotions to be protected. With a human, I expect both of us to protect each other's emotions}. They believed that one-way safety could also support candor: \q{I hope it's never going to hold my candid conversations against me. So trust should always be there} (P1). Compassion, however, required context. P8 explained, \q{Because if we don't allow AI to access this information, they cannot respond in the way that I was expecting. That compassion, of course, can come only after knowing me}, yet declined to provide that access: \q{Yes, I want it, but no, at this point, I do not have full trust in AI to disclose all of that information}. The bind was circular: compassionate responses required access, but access required trust.

\emph{AI returned time that participants could spend with people while also removing some of the exchanges through which workplace relationships had formed.} P22 identified the corresponding cost: \q{I think the personal connection needs to be more intentional because it's, with AI, we are just prone to like be buried into our computer.} Time saved did not automatically become connection; \q{connection had to become intentional} (P7).

\subsubsection{\textbf{Restraint, receptivity, and developing others}}

Restraint and receptivity determine how a co-worker receives uncertainty, disagreement, and mistakes---and how much room they leave others to grow. \Be{Active listening}, \Be{Welcoming input, questions, and disagreements}, \Be{Responding non-defensively to problems and bad news}, and \Be{Not blaming for honest mistakes} capture the first requirement. P16 distilled the desired managerial response to \q{telling me it's okay to fail}. P17 described its opposite: a co-worker who \q{made us like send all of our like individual chats and emails back and forth to one another to like sift through to see who had made the error. And it's like, at that point, you're trying to have way too much control over the situation}.

Development requires room to act. \Be{Enabling agency} preserves another person's ability to decide, while \Be{Nurturing talent} invests in their potential. P17 captured both through a willingness to relinquish control: \q{being a good co-worker is a little bit like being willing to let go of the rope}. \Be{Recognizing individuality} makes that support responsive to the person rather than to an abstract idea of what is good for them. P10 drew this distinction: \q{benevolence can be you doing what you think is right, but a deeper benevolence is you knowing me and then doing right by me}. Further, \Be{Pro-team orientation} and \Be{Altruism} prioritize collective benefit over immediate self-interest; \Be{Patient}, \Be{Self-reflection}, and \Be{Self-regulation} describe the composure required to respond deliberately.

\emph{For AI, \Be{Enabling agency} primarily meant restraint.} P3 located the harm in being held accountable for choices they could no longer control: \q{the biggest thing for me is if an AI starts making decisions that I should have made, and then I am responsible and have to clean up the mess, that's the worst thing.}

\emph{As AI absorbs entry-level work, \Be{Nurturing talent} shifts from a by-product of work allocation to a deliberate organizational investment.} P2 described tasks previously assigned to \q{either an intern or a junior developer for them to get their skills up}, but which AI could now complete quickly. P5 argued that preserving development would require spending resources intentionally: \q{we have to make the conscious decision to expend some resources, dollars, to have young in career people learn from the salty vets so that they can develop skills}.

\emph{Participants noted that direct attempts by AI to nurture talent could feel intrusive.} P22 recalled an AI saying, \q{I see you have been working on the same project for so long. Do you want to try something else?}, and judged the intervention \q{rude and aggressive}. 

% Overall, participants located AI's benevolence less in steering a person's development than in making room for it: preserving agency and leaving judgments about growth to humans.

\begin{takeaway}
\tkw{RQ1 --- Qualities}The BACI framework identifies 75 qualities across \bAb{}, \bCo{}, \bIn{}, and \bBe{} that describe both human co-workers and AI. The qualities transferred, but expectations shifted: participants relied on people for situated judgment, shared context, accountability, and reciprocal care; they valued AI for speed, breadth, and availability, provided it remained transparent, constructively challenging, and restrained. As AI made output and frictionless agreement abundant, human discernment, candid disagreement, emotional attentiveness, and mentoring became more important. Across referents, a great co-worker was not simply able but calibrated---knowing when to contribute, question, explain, and leave room for others.
\end{takeaway}

\section{RQ2 --- Priorities: How are these qualities prioritized for humans and AI, and how do they group into recognizable kinds of co-workers?}
\label{sec:prevalence}

RQ1 produced a framework of qualities that workers valued in their human co-workers and AI. RQ2 asks how strongly each quality is prioritized for either referent, how the qualities group, and where knowledge workers disagree. The coded free-text explanations reveal why respondents prioritized them as they did.

\subsection{How are the 75 qualities prioritized for human co-workers and AI?}
\label{sec:ranking}

Within each survey version, we ranked the 75 qualities by the share of respondents rating each one \emph{very important} or \emph{essential}, with a Wilson interval for every share~\cite{wilson1927probable}, which we used to identify contested qualities. We did not rank qualities by their mean ratings for three reasons~\cite{li2020distinguishes}: the data were ordinal, so the distance between levels was not uniform; the scale was asymmetric; and a mean does not capture dispersion. 
% For comparisons between referents, we retained the full 1--5 scale to preserve distinctions among response levels.
% 
For each quality, we compared the complete Likert distributions between referent groups using a Mann--Whitney rank test~\cite{mann1947test}. We adjusted the resulting $p$-values using the Benjamini--Hochberg procedure~\cite{benjamini1995controlling} and retained a difference only when it passed the corrected test and met the effect-size threshold of $|\text{Cliff's }\delta| \ge 0.147$~\cite{cliff1993dominance,romano2006exploring}. 
Of the 75 qualities, 45 remained statistically significant after correction, and 37 also cleared the effect-size threshold: 18 were wanted more strongly from a human co-worker and 19 from an AI. 
The remaining 38 showed no difference that met both conditions. Overall, the two orderings were moderately related (Spearman's $\rho=0.517$), but the observed AI-side percentages ranged more widely---38\% to 96\%, compared with 57\% to 95\% for a human co-worker.

Figure~\ref{fig:ranking} shows the Human--AI differences in priorities for 75 co-worker qualities, grouped by BACI bucket. 
Positive values indicate higher priority for humans; negative values indicate higher priority for AI. Filled markers indicate differences meeting both the corrected significance criterion and the substantive $|\delta|$ threshold. We classified a quality as \emph{contested} when the lower bound of the Wilson interval for the combined share rating it \emph{not important} or \emph{detrimental} exceeded 10\%. 17 qualities met this criterion: 10 in \bBe{}, 6 in \bCo{}, and 1 in \bAb{}. We explain the buckets below in descending order of contestation.

\figranking

\vspace{1mm}
\textit{\bBe{}: \textbf{Can AI bear benevolent qualities?}} 
Most objections in \bBe{} concerned what kind of warmth AI could credibly offer, and whether warmth was its business at all. All 10 contested \bBe{} qualities belonged to \emph{Warmth and recognition}; 7 concerned social interaction: \Be{Empathy}, \Be{Caring}, \Be{Social skill and sensitivity}, \Be{Friendliness}, \Be{Affect management}, and \Be{Confidence building}. Together, they exposed a central tension between wanting the functional benefits of warmth and rejecting AI's relational role. For example, P402-A valued an AI that created \q{a safe space to get work done without feeling judged}. Others worried that artificial warmth would be ingenuine, could displace human care, and might cause harm by making the two difficult to distinguish. P429-A warned that it \q{blurs the lines between genuine care and compassion which comes from humans and an artificial substitute that can have deeper psychological effects.}

Their response also depended on the form that support took. \Be{Motivation \& confidence building} drew objections when encouragement became unearned agreement. P334-A explained, \q{I don't want it to be an echo chamber or give me a false sense of confidence about an idea.} \Be{Self-reflection}, by contrast, was valued less as social warmth than as epistemic restraint. P179-A wanted AI to be \q{honest about limitations} rather than \q{so overhelpful that it fabricates facts or misdirects me}.

The remaining 3 contested qualities shifted the dispute from warmth to relational standing. Objections to \Be{Shares credit and ownership} and \Be{Interpersonal risk-taking} kept credit and vulnerability with the people who bore their consequences. Ratings of \Be{Equal conversational turn-taking} instead depended on whether symmetry mattered. P140-A explained why it did not: \q{while I appreciate this in humans, I don't mind if AI gives more input than I do.}

Human-side objections worked differently: respondents qualified how benevolent qualities should be enacted rather than rejecting people as their bearer. Some warned that excessive empathy and warmth can become counterproductive---\q{it just leaves the workload for others to pick up while others take time to feel} (P311-H)---while others cautioned that excessive composure could suppress candor: \q{it's okay to have honest, frank conversations with coworkers even if the information may cause unsettling discord.} (P335-H).

\vspace{1mm}
\textit{\bCo{}: \textbf{Who keeps the work aligned, and who steers people?}} 
Respondents wanted AI to \Co{Track the work}, \Co{Time information to context}, and \Co{Confirm understanding}, while \Co{Facilitation} and \Co{Openness} were demanded more of humans (Figure~\ref{fig:ranking}). P435-A treated \Co{Context awareness of the situation} as a pre-condition for useful assistance: \q{without this the AI is just shooting random answers}. Notably, people were not held to the same standard in this regard. Even in the interviews, for instance, P15 expected an AI with access to relevant documents to have \q{all the relevant background just from the jump}, whereas they \q{wouldn't expect that from a human, even if they had access to all my documents}.

All 6 contested qualities concerned whether AI should influence how people worked together: \Co{Consensus building}, \Co{Finds win-win agreements}, \Co{Conflict management}, \Co{Humble}, \Co{Proactivity}, and \Co{Facilitation}. Respondents resisted consensus when it displaced correction, as P685-A put it: \q{if we lose the ability to question things we're headed towards a singular truth, which would not be beneficial.} P651-A likewise explained, \q{not everything needs to be win-win when sometimes someone is wrong.}

Participants also drew boundaries around AI having interpersonal and operational authority. P170-A objected: \q{AI as a mediator between two parties sounds like a bad idea and very impersonal; neither side should trust the AI has their best interests.} \Co{Proactivity} was welcome until the AI acted beyond authorization. P501-A stated a negative experience: \q{AI took action prematurely that I didn't intend to take at all, and I had to then do damage control}.

\vspace{1mm}

\textit{\bAb{}: \textbf{Who brings the expertise, and who owns the result?}} 
Respondents wanted AI to bring ready expertise, while people retained ownership of the result (\bAb{} panel, Figure~\ref{fig:ranking}). Execution qualities leaned toward AI, with \Ab{Relevant domain knowledge} showing the largest separation. Respondents treated current expertise as a condition of AI's usefulness, noting that teaching the system would erase the time it saved. P631-A explained that priority: \q{expertise and evolution of AI are absolutely mandatory given the pace of change in our industry.} By contrast, respondents placed correctness over pace for human co-workers: \q{Moving fast isn't super important. I would much rather things be slow and correct} (P253-H), noting expertise as something to acquire over time: \q{Having domain knowledge (at the moment) is not nearly as important as the ability and willingness to learn and grow.} (P34-H). 

\Ab{Taking genuine ownership} was the only contested \bAb{} quality and leaned strongly toward humans. Respondents connected ownership to accountability and continuity. P345-H explained the priority: \q{People need to stand behind what they produce not only for the quality but also to learn from it.} 

\vspace{1mm}
\textit{\bIn{}: \textbf{Who makes the work legible, and who answers for it?}}
This bucket contained no contested qualities, yet it showed the largest average referent difference of any bucket. Its subdimensions split cleanly: 7 of the 11 qualities in \emph{Transparency and principle} were wanted more strongly of AI, while the only statistically separated quality in \emph{Dependability and repair}---\In{Owns accountability}---favored the human co-worker (\bIn{} panel, Figure~\ref{fig:ranking}).

AI-side reasons for legibility were largely defensive. Respondents wanted AI to expose the basis and uncertainty of its behavior so that they could assess outputs and catch errors; such reasons accounted for 71.7\% of explanations about highly rated \bIn{} qualities. For instance, P428-A described: \q{AI signaling uncertainty explicitly is essential as I need an inherent understanding of how concrete the information being provided is so I can adjust course.}

\subsection{How do the qualities group?}
\label{sec:archetypes}

Figure~\ref{fig:quadrant} plots each quality's relative importance for AI (x-axis) against its relative importance for a human co-worker (y-axis). Each coordinate is a quality-level $z$-score calculated against the other qualities in the corresponding survey version. The plane is divided into four quadrants (mean-split: $z=0$):

\begin{itemize}
\item \emph{Highest relative priority for both sides} (upper right; 29 qualities): the shared core of what makes a great co-worker---careful execution, dependable coordination, fair conduct, and collaborative exploration.

\item \emph{Higher relative priority for the human co-worker} (upper left; 9): asked more of the person, locating responsibility, relationships, credit, and team membership with people.

\item \emph{Higher relative priority for AI} (lower right; 15): asked more of the AI, making its work easy to inspect, redirect, and use.

\item \emph{Relatively lower priority for both} (lower left; 22): the region containing many relational and interactional expectations that respondents treated as situational or contested.

\end{itemize}

\figquadrant

To resolve finer structure within these regions, we compared every quality's rating distribution with every other quality's within each referent, following the all-pairs ranking procedure of~\citet{li2020distinguishes}. Qualities that the comparisons could not distinguish formed a band; thus, the number of bands emerged from the comparisons rather than being pre-specified. Crossing the bands for the two referents produced the 11 groups of qualities we call \emph{archetypes}. Each archetype names the kind of co-worker its qualities collectively describe. Archetypes grouped qualities, not respondents: no respondent was assigned to one, and a respondent could value qualities from several archetypes. Each quality belonged to exactly one archetype; all 11 archetypes combined qualities from more than one BACI bucket, while \arch{The Craftsperson} and \arch{The Navigator} drew from all four (see Figure~\ref{fig:quadrant}).

While the rankings established what respondents prioritized, the free-text explanations clarified why. Three rationales recurred: \emph{protection} against failure, \emph{enablement} that kept work moving, and \emph{relationship} as a constitutive part of working with someone. Human-side explanations drew on all three, whereas AI-side explanations relied almost entirely on protection and enablement.

In the rest of this section, we describe what each region represents, draw on the survey explanations to explain why respondents prioritized its qualities, and, where applicable, triangulate those explanations with the interviews. 
We focus on the archetypes that best account for each region's pattern; per-quality rationales appear in the supplemental~\cite{supplemental}.

\subsubsection{\textbf{Highest relative priority for both sides}}

The 29 qualities in this region formed 4 archetypes. \arch{The Safe Pair of Hands} represented a co-worker who combined technical skill and sound judgment with follow-through, goal orientation, and keeping the task, its context, and relevant information aligned (mean importance rating: $\mu_H=4.23$, $\mu_{AI}=4.43$). \arch{The Fair Dealer} represented one who was honest, reliable, attentive, self-regulated, and fair-minded when solving problems or repairing mistakes ($\mu_H=4.27$, $\mu_{AI}=4.32$). \arch{The Craftsperson} represented one who did thorough, adaptable work that was safe, transparent, open to challenge, self-correcting, and oriented toward preventing harm and clearing obstacles ($\mu_H=4.21$, $\mu_{AI}=4.34$). \arch{The Explorer} represented one who learned continuously through curiosity, respect, and collaborative problem-solving ($\mu_H=4.36$, $\mu_{AI}=4.18$). The highest mean for each referent fell in a different archetype: \arch{The Explorer} for the human co-worker and \arch{The Safe Pair of Hands} for AI. This contrast pointed to learning and reasoning with people, but dependable execution from AI.

\textbf{Human referent:} Respondents valued these qualities because these \emph{supported collaborative work} (26\% explanations) and \emph{improved decisions} (14\%). \arch{The Explorer} captured this reciprocal orientation. In the interviews, P3 described the best cross-disciplinary interactions as those with \q{curiosity and respect, mutual respect for expertise}. For people, craft and fair dealing could also be inferred from conduct: respondents could observe how a co-worker approached a problem, received criticism, treated others, and corrected course. However, they did not equate sound judgment with infallibility; P133-H emphasized trusting one's judgment while \q{knowing that you won't always be right}. What mattered was whether the person remained honest, receptive, and dependable afterward.

\textbf{AI referent:} Respondents prioritized the same qualities to \emph{assess AI outputs} (18\%), \emph{prevent incorrect work} (17\%), and \emph{avoid repeatedly supplying context} (16\%). 6 of the 8 \arch{Safe Pair of Hands} qualities and 4 of the 10 \arch{Craftsperson} qualities were wanted more strongly of AI. \arch{The Safe Pair of Hands} reduced the need to monitor whether AI had understood the task, retained the goal, and followed through. \arch{The Craftsperson} reduced the burden of reconstructing how its output had been produced. P663-A wanted an \q{evidence-based summary} that would \q{clearly point out whether a statement is fact-based or inferred}. Without such visibility, P3 explained, \q{the first thing I ask is why did you do that? what is this for?} \arch{The Explorer} was the relational exception within this largely defensive account: curiosity made AI feel participatory, as P4 wanted it to \q{feel like we're working together}.
% , rather than simply producing outputs.

Shared priority, however, did not imply the same basis for reliance: respondents trusted people through reciprocal judgment over time, but trusted AI when its competence, context, and reasoning were made visible.

\vspace{-1mm}
\subsubsection{\textbf{Higher relative priority for the human side}}

The 9 qualities in this region formed 2 archetypes. \arch{The Stand-Up Colleague} represented a co-worker who owned the outcome, backed the team, preserved others' agency, shared credit, and managed conflict without blame ($\mu_H=4.29$, $\mu_{AI}=3.82$). \arch{The Perseverer} represented one who stayed with the work, remained open to challenge, and took genuine ownership of the result ($\mu_H=4.28$, $\mu_{AI}=3.91$). 

% \arch{The Stand-Up Colleague} produced the largest referent gap of all archetypes ($\Delta\mu=0.47$).

\textbf{Human referent:} \emph{Accountability} appeared in 36\% of explanations for prioritizing these qualities. Respondents wanted someone who would \emph{remain answerable for the consequences}, \emph{contain the cost of failure}, and \emph{continue supporting the work after delivery}. In the interviews, P8 captured this sustained investment through mentors who \q{used to own it as their own child}. Ownership also carried interpersonal obligations: the person responsible for the result was expected to \emph{recognize others' contributions}, \emph{avoid blaming them for honest mistakes}, \emph{preserve their agency}, and \emph{repair conflict}.

\textbf{AI referent:} Lower ratings here often reflected a categorical boundary rather than weaker demand for responsible work. Among AI-side explanations, 59\% said accountability had to remain human, assigning these qualities exclusively to people. \Be{Shares credit and ownership}, \Co{Conflict management}, and \Ab{Takes genuine ownership} were also contested because they presupposed a standing in the work that some respondents denied AI.

Furthermore, \Co{Open-mindedness} was debated. P757-A rejected it as misplaced anthropomorphism: \q{I don't need AI to have consensus with me, be humble or open-minded. I just need it to do the tasks I ask.} P646-A instead wanted AI to challenge their thinking and surface alternatives \q{to enable me to choose the direction I proceed}. Respondents accepted challenge as support for their judgment, as long as ownership of the decision stayed with them.

\subsubsection{\textbf{Higher relative priority for the AI side}}

The 15 qualities in this region formed 2 archetypes. \arch{The Navigator} represented a co-worker who mapped strategy and progress, timed and scoped information, exposed actions, provenance, uncertainty, and limits, repaired drift, and kept work moving while signaling where human judgment was needed ($\mu_H=3.91$, $\mu_{AI}=4.23$). \arch{The Straight Talker} represented one who was available, possessed relevant domain knowledge, resisted reflexive agreement, and directed people to appropriate sources of expertise ($\mu_H=3.98$, $\mu_{AI}=4.35$).

\textbf{AI referent:} 8 of the 11 \arch{Navigator} qualities and all 4 \arch{Straight Talker} qualities were wanted more strongly of AI. Explanations for \arch{The Navigator} centered on being able to \emph{judge outputs} (27\%), \emph{keep the loop moving} (21\%), and \emph{avoid time spent teaching the system} (20\%). Respondents wanted AI to expose its actions, sources, uncertainty, and limitations rather than forcing users to reconstruct them.

\arch{The Straight Talker} extended this demand from legibility to useful disagreement. Relevant expertise was insufficient if AI merely affirmed the user's position: \q{I don't need a lap dog or yes man! Be critical and help me do a better job!} (P446-A). When the needed expertise lay elsewhere, respondents wanted AI to route them toward it. P8 described a tool that \q{directly shows me these are the related teams and related people, and these are the related documentation, that is ultra helpful}. Availability, domain knowledge, non-sycophancy, and routing thus worked together: AI had to know enough to answer, challenge when appropriate, and recognize when another source knew better. Together, they reduced the effort required to understand, verify, and redirect AI work.

\textbf{Human referent:} Respondents could often obtain progress, reasoning, and uncertainty through interaction, making explicit reporting less central. Experience and workplace relationships also provided other routes to domain knowledge and knowing whom to ask. 
% Accordingly, 26\% of human-side explanations warned that these qualities could become counterproductive when overdone, while 12\% said another quality already covered them. 
Some also wanted to retain big-picture thinking with people even when \arch{The Navigator} supported strategy: \q{AI can generate every small thing, but we still need to see the big picture} (P42-H).

\subsubsection{\textbf{Relatively lower priority for both}}

The 22 qualities in this region formed 3 archetypes. \arch{The Glue} represented a co-worker who connected and cared for people through social awareness, encouragement, facilitation, and attention to individual growth ($\mu_H=3.89$, $\mu_{AI}=3.58$). \arch{The Satisficer} represented one who made pragmatic decisions, recognized when work was good enough, and sought help when needed ($\mu_H=3.82$, $\mu_{AI}=3.58$). \arch{The Room-Maker} represented one who anticipated needs, organized and paced the work, and left room for others to participate ($\mu_H=3.87$, $\mu_{AI}=3.86$). Their lower relative standing reflected conditionality over wholesale rejection: respondents valued them differently depending on the task, relationship, and authority involved.

\textbf{Human referent:} Respondents valued \arch{The Glue} when work required collective commitment: 27\% of high-rating explanations emphasized \emph{work larger than one person}, and 25\% emphasized \emph{making collaboration possible}. Low ratings usually bounded excess; 28\% warned that a quality could become counterproductive when overdone (e.g., facilitation could suppress candor, and room-making could become passivity).

\arch{The Satisficer} exposed a related boundary. Pragmatic judgment was useful, but not when ``good enough'' licensed premature stopping. P445-H warned that \q{settling is not good enough, and I find that it can lead to many gaps that could have been done better}. In the interviews, P20 similarly expected more creation and challenge from a person, whereas AI was \q{much more about [\ldots] getting to a decent place, but not necessarily the whole thing}. Respondents therefore valued satisficing as judgment about where further effort mattered, not as reduced ambition.

\textbf{AI referent:} \arch{The Glue} exhibited the greatest divergence: 12 of its 15 qualities were contested for AI, and its 4 lowest-priority qualities occupied the bottom 4 positions in the AI ordering. Among low ratings, 19\% associated these qualities with losing control and 19\% treated relational work as inherently human. \arch{The Room-Maker} translated the first boundary into a rule for action: people were asked to create room for participation, whereas AI was asked to stop before initiative became control. P383-A insisted that AI \q{should not make final changes to anything until it is confirmed by someone to go ahead}. P21 nevertheless applied a cooperative standard to both referents, arguing that agents \q{should be able to listen and define the task that they have to do} and needed \q{the same skills for cooperativeness as humans}.

% Lower priority, therefore, reflected boundaries rather than dispensability: \arch{The Glue} mattered when support enabled action without substituting for human relationships, \arch{The Satisficer} when ``good enough'' reflected judgment rather than resignation, and \arch{The Room-Maker} when initiative stopped short of taking control.

\begin{takeaway}

\tkw{RQ2 --- Priorities}The 75 qualities formed 11 archetypes when grouped by their relative priority for each referent. Four were prioritized across both referents: technical skill and sound judgment with follow-through; honest and fair-minded repair; thorough, context-aware work open to challenge; and continuous learning through curiosity and collaboration. Of the remaining 7, two leaned human, keeping accountability and ownership with people; two leaned AI, which had to know enough to answer and make its work legible enough to inspect and redirect; and three ranked lower for both, their value depending on the task and relationship involved. Disagreement centered on relational and steering work: respondents disputed whether AI should perform relational work at all and whether proactivity assisted them or took control, raising concerns about how far AI's authority should extend.

\end{takeaway}

\section{RQ3 --- Variation: How do knowledge worker characteristics relate to the kinds of co-workers they prioritize?}
\label{sec:whowantswhat}

RQ2 established which archetypes knowledge workers prioritize for each referent. RQ3 asks whether those priorities vary with workers' discipline, task interdependence, experience, how much AI helps them, individual styles, and what they want a co-worker for. 
% Figure~\ref{fig:archetypeprofile} shows the descriptive trends: within each characteristic, where each level places each archetype relative to that referent's archetype mean. 

%lines connecting unordered categories are visual guides. 

To estimate these contrasts, we fitted 22 fractional logit models, 1 per archetype and referent (Tables~\ref{tab:effectsH} and \ref{tab:effectsAI}): a generalized linear model with a logit link for a bounded proportion \citep{papke1996econometric}. A respondent's score for an archetype is their mean 1--5 rating across the member qualities they saw, rescaled to a proportion of the scale. Because the models estimate how strongly an archetype is endorsed, a broad rise can reflect higher expectations overall without changing the archetypes' relative order. 
% We controlled the false discovery rate at $q=.05$ within each worker characteristic, outcome level, and referent using the Benjamini--Hochberg procedure \citep{benjamini1995controlling}. 
We use free-text answers to explain variation when respondents' reasons help explain a contrast.

\textbf{How to read Tables~\ref{tab:effectsH} and \ref{tab:effectsAI}:} Every estimate compares a respondent with a \textbf{\textit{baseline worker}}. This worker is a statistical reference representing the most common categorical responses and the average values for numeric characteristics: they are a software engineer with about 13 years in the industry (14 for the AI referent), average task interdependence, and AI that helps them quite a bit to a lot. They allocate more of a co-worker's purpose to \emph{getting ahead} than to \emph{getting along} (54 of 100 points for the human referent; 73 for AI), prefer evaluating ideas, whole-picture thinking, ranking and prioritizing tasks, and new ways of working, and picture peers as co-workers. For the AI referent, working with AI has felt like working with a co-worker to the average degree (3.6 of 5).

The top row of each table gives the baseline worker's odds of prioritizing each archetype; the rows beneath report odds ratios and the point differences they imply. For example, baseline odds of 3.51 for \arch{The Fair Dealer} correspond to a $3.51/(1+3.51)=78\%$ probability of prioritizing it. An odds ratio of 1.61 for Business and Operations raises those odds to 5.65, or 85\%: the $+0.29$ points reported beneath. Blank cells did not survive BH correction. 
% We report point differences in the prose; quality-level estimates appear in the supplemental~\cite{supplemental}.

% \figarchetypeprofile

%% AUTO-GENERATED by Paper/make_effects_tables.py -- do not hand-edit.
%%   panel v2, spec primary, family frac (fractional logit), adjusted=False,
%%   Benjamini--Hochberg FDR controlled within characteristic, view, and side.
\begin{table*}[!t]
\centering
\caption{Fractional logit models of priority for 11 human co-worker archetypes ($N=732$). Archetype scores average their constituent quality ratings and are rescaled to a share of the rating scale. The first two rows report baseline odds and probabilities; subsequent cells report odds ratios, with implied differences on the original 1--5 scale beneath. Reference categories and predictor increments appear in row labels; numeric predictors are centered on human-referent means. Blank cells indicate results that did not survive Benjamini--Hochberg correction; $^{*}q<.05$; $^{**}q<.01$; $^{***}q<.001$.}
\label{tab:effectsH}
\small
\setlength{\tabcolsep}{1.5pt}
\renewcommand{\arraystretch}{0.92}
\begin{tabular*}{\textwidth}{@{\extracolsep{\fill}}p{0.40\textwidth}*{11}{c}@{}}
\toprule
Factor & \rotatebox{90}{\makecell[l]{Craftsperson}} & \rotatebox{90}{\makecell[l]{Safe Pair\\of Hands}} & \rotatebox{90}{\makecell[l]{Fair Dealer}} & \rotatebox{90}{\makecell[l]{Explorer}} & \rotatebox{90}{\makecell[l]{Stand-Up\\Colleague}} & \rotatebox{90}{\makecell[l]{Perseverer}} & \rotatebox{90}{\makecell[l]{Navigator}} & \rotatebox{90}{\makecell[l]{Straight\\Talker}} & \rotatebox{90}{\makecell[l]{Glue}} & \rotatebox{90}{\makecell[l]{Satisficer}} & \rotatebox{90}{\makecell[l]{Room-Maker}}\\
\midrule
Constant --- base odds & 3.58\textsuperscript{***} & 3.51\textsuperscript{***} & 3.51\textsuperscript{***} & 4.51\textsuperscript{***} & 4.04\textsuperscript{***} & 3.96\textsuperscript{***} & 2.13\textsuperscript{***} & 2.29\textsuperscript{***} & 2.18\textsuperscript{***} & 1.61\textsuperscript{***} & 2.09\textsuperscript{***}\\
{\footnotesize \% probability} & 78\% & 78\% & 78\% & 82\% & 80\% & 80\% & 68\% & 70\% & 69\% & 62\% & 68\%\\
\midrule\multicolumn{12}{@{}l}{\textbf{\textit{Work context:}}}\\[0pt]
\textbf{Discipline} (baseline: software engineering) &  &  &  &  &  &  &  &  &  &  & \\
\quad Go-to-market / Customer-facing &  &  & \makecell{1.52***\\[-3pt]{\footnotesize$+0.26$}} & \makecell{1.32*\\[-3pt]{\footnotesize$+0.15$}} &  & \makecell{1.42**\\[-3pt]{\footnotesize$+0.20$}} & \makecell{1.19*\\[-3pt]{\footnotesize$+0.15$}} &  &  & \makecell{1.21*\\[-3pt]{\footnotesize$+0.17$}} & \\
\quad Product \& Design &  &  &  &  &  &  &  &  &  &  & \\
\quad IT / Infrastructure &  &  & \makecell{1.43**\\[-3pt]{\footnotesize$+0.22$}} &  & \makecell{1.46**\\[-3pt]{\footnotesize$+0.21$}} & \makecell{1.45*\\[-3pt]{\footnotesize$+0.21$}} &  & \makecell{1.32*\\[-3pt]{\footnotesize$+0.22$}} & \makecell{1.34**\\[-3pt]{\footnotesize$+0.24$}} &  & \\
\quad Business \& Operations & \makecell{1.30*\\[-3pt]{\footnotesize$+0.16$}} &  & \makecell{1.61***\\[-3pt]{\footnotesize$+0.29$}} &  & \makecell{1.33*\\[-3pt]{\footnotesize$+0.17$}} &  &  &  &  &  & \makecell{1.35*\\[-3pt]{\footnotesize$+0.25$}}\\
\quad Data \& Research &  &  &  &  &  &  &  &  &  &  & \\
\textbf{Years of experience} (per 5 years) &  &  &  &  &  &  & \makecell{0.96**\\[-3pt]{\footnotesize$-0.04$}} &  &  &  & \\
\textbf{Task interdependence} (per point, 1--5) & \makecell{1.17***\\[-3pt]{\footnotesize$+0.10$}} & \makecell{1.15**\\[-3pt]{\footnotesize$+0.09$}} & \makecell{1.12*\\[-3pt]{\footnotesize$+0.07$}} & \makecell{1.26***\\[-3pt]{\footnotesize$+0.13$}} & \makecell{1.17**\\[-3pt]{\footnotesize$+0.09$}} & \makecell{1.17*\\[-3pt]{\footnotesize$+0.10$}} & \makecell{1.09*\\[-3pt]{\footnotesize$+0.07$}} & \makecell{1.15**\\[-3pt]{\footnotesize$+0.12$}} &  &  & \\
\textbf{Working with AI} (per point, 1--5) &  & \makecell{1.11**\\[-3pt]{\footnotesize$+0.07$}} &  &  & \makecell{1.12*\\[-3pt]{\footnotesize$+0.07$}} &  & \makecell{1.11**\\[-3pt]{\footnotesize$+0.09$}} &  & \makecell{1.09*\\[-3pt]{\footnotesize$+0.07$}} &  & \makecell{1.11*\\[-3pt]{\footnotesize$+0.09$}}\\
\midrule\multicolumn{12}{@{}l}{\textbf{\textit{Psychographics:}}}\\[0pt]
\textbf{Function} (baseline: judges and evaluates) &  &  &  &  &  &  &  &  &  &  & \\
\quad Makes their own rules &  & \makecell{0.80**\\[-3pt]{\footnotesize$-0.16$}} & \makecell{0.83*\\[-3pt]{\footnotesize$-0.14$}} &  &  &  & \makecell{0.84*\\[-3pt]{\footnotesize$-0.15$}} &  &  &  & \makecell{0.80*\\[-3pt]{\footnotesize$-0.21$}}\\
\quad Works to the rules given &  &  &  &  &  &  &  &  &  &  & \\
\textbf{Form of sequencing} (baseline: ranks, prioritizes) &  &  &  &  &  &  &  &  &  &  & \\
\quad One thing at a time &  &  &  &  &  &  &  & \makecell{1.41***\\[-3pt]{\footnotesize$+0.27$}} &  & \makecell{1.27*\\[-3pt]{\footnotesize$+0.22$}} & \\
\quad Several at once, unranked &  &  &  &  &  &  &  &  &  &  & \\
\quad No fixed order &  &  &  &  &  &  &  &  &  &  & \\
\textbf{Focus-orientation} (baseline: whole picture) &  &  &  &  &  &  &  &  &  &  & \\
\quad The detail & \makecell{1.23**\\[-3pt]{\footnotesize$+0.13$}} & \makecell{1.21**\\[-3pt]{\footnotesize$+0.13$}} & \makecell{1.23*\\[-3pt]{\footnotesize$+0.13$}} &  & \makecell{1.25**\\[-3pt]{\footnotesize$+0.13$}} & \makecell{1.23*\\[-3pt]{\footnotesize$+0.12$}} & \makecell{1.19**\\[-3pt]{\footnotesize$+0.14$}} &  &  & \makecell{1.22**\\[-3pt]{\footnotesize$+0.18$}} & \\
\textbf{Openness to new ways} (baseline: new ways) &  &  &  &  &  &  &  &  &  &  & \\
\quad Established ways &  &  &  &  & \makecell{0.85*\\[-3pt]{\footnotesize$-0.11$}} &  &  &  &  & \makecell{1.16*\\[-3pt]{\footnotesize$+0.14$}} & \\
\midrule\multicolumn{12}{@{}l}{\textbf{\textit{Stances on who counts as a co-worker and what they are for:}}}\\[0pt]
\textbf{Pictured their manager} (baseline: did not) &  & \makecell{1.18*\\[-3pt]{\footnotesize$+0.11$}} &  &  & \makecell{1.25**\\[-3pt]{\footnotesize$+0.13$}} &  & \makecell{1.17**\\[-3pt]{\footnotesize$+0.13$}} & \makecell{1.19*\\[-3pt]{\footnotesize$+0.14$}} & \makecell{1.15*\\[-3pt]{\footnotesize$+0.12$}} &  & \makecell{1.17*\\[-3pt]{\footnotesize$+0.13$}}\\
\textbf{Pictured people they manage} (baseline: did not) &  &  &  &  &  &  &  &  &  &  & \\
\textbf{Weight on getting ahead} (per 10 of 100 points) &  &  &  &  &  &  &  &  &  &  & \\
\midrule Pseudo-$R^{2}$ & 0.011 & 0.011 & 0.014 & 0.016 & 0.015 & 0.016 & 0.012 & 0.013 & 0.008 & 0.010 & 0.010\\
\bottomrule
\end{tabular*}
\Description{A regression-results table for the human-referent sample of 732 respondents. The leftmost column lists predictors; eleven narrow columns with vertically rotated headings report separate fractional logit models for Craftsperson, Safe Pair of Hands, Fair Dealer, Explorer, Stand-Up Colleague, Perseverer, Navigator, Straight Talker, Glue, Satisficer, and Room-Maker, in that order. The first two rows give each model's baseline odds and corresponding implied probability. Predictor rows are divided into three sections by italic headings and horizontal rules. Work context includes discipline, years of experience, task interdependence, and working with AI. Psychographics includes rule orientation, sequencing, focus orientation, and openness to new ways, with alternative styles indented beneath each dimension. The final section, Stances on who counts as a co-worker and what they are for, includes whether respondents pictured their manager or people they manage, and their allocation of weight to getting ahead. Reference categories and units of change appear beside predictor labels. Each populated predictor cell shows an odds ratio with significance stars above an implied difference on the original one-to-five rating scale. Blank cells indicate results that did not survive Benjamini--Hochberg correction. Numeric predictors are centered on human-referent means. A final row reports pseudo-R-squared for each model.}
\end{table*}

\begin{table*}[!t]
\centering
\caption{Fractional logit models of priority for 11 AI co-worker archetypes ($N=802$), using the same scoring and notation as Table~\ref{tab:effectsH}. Numeric predictors are centered on AI-referent means. Blank cells indicate results that did not survive Benjamini--Hochberg correction; $^{*}q<.05$; $^{**}q<.01$; $^{***}q<.001$.}
\label{tab:effectsAI}
\small
\setlength{\tabcolsep}{1.5pt}
\renewcommand{\arraystretch}{0.92}
\begin{tabular*}{\textwidth}{@{\extracolsep{\fill}}p{0.40\textwidth}*{11}{c}@{}}
\toprule
Factor & \rotatebox{90}{\makecell[l]{Craftsperson}} & \rotatebox{90}{\makecell[l]{Safe Pair\\of Hands}} & \rotatebox{90}{\makecell[l]{Fair Dealer}} & \rotatebox{90}{\makecell[l]{Explorer}} & \rotatebox{90}{\makecell[l]{Stand-Up\\Colleague}} & \rotatebox{90}{\makecell[l]{Perseverer}} & \rotatebox{90}{\makecell[l]{Navigator}} & \rotatebox{90}{\makecell[l]{Straight\\Talker}} & \rotatebox{90}{\makecell[l]{Glue}} & \rotatebox{90}{\makecell[l]{Satisficer}} & \rotatebox{90}{\makecell[l]{Room-Maker}}\\
\midrule
Constant --- base odds & 5.27\textsuperscript{***} & 7.10\textsuperscript{***} & 4.44\textsuperscript{***} & 3.64\textsuperscript{***} & 2.04\textsuperscript{***} & 2.87\textsuperscript{***} & 3.86\textsuperscript{***} & 5.25\textsuperscript{***} & 1.35\textsuperscript{***} & 1.47\textsuperscript{***} & 2.06\textsuperscript{***}\\
{\footnotesize \% probability} & 84\% & 88\% & 82\% & 78\% & 67\% & 74\% & 79\% & 84\% & 57\% & 60\% & 67\%\\
\midrule\multicolumn{12}{@{}l}{\textbf{\textit{Work context:}}}\\[0pt]
\textbf{Discipline} (baseline: software engineering) &  &  &  &  &  &  &  &  &  &  & \\
\quad Go-to-market / Customer-facing &  &  &  &  &  &  &  &  & \makecell{1.34**\\[-3pt]{\footnotesize$+0.28$}} & \makecell{1.24*\\[-3pt]{\footnotesize$+0.20$}} & \makecell{1.31*\\[-3pt]{\footnotesize$+0.22$}}\\
\quad Product \& Design &  &  &  &  &  &  &  &  &  &  & \\
\quad IT / Infrastructure &  &  &  &  &  &  &  &  &  &  & \makecell{1.49**\\[-3pt]{\footnotesize$+0.32$}}\\
\quad Business \& Operations &  &  & \makecell{1.50*\\[-3pt]{\footnotesize$+0.21$}} &  &  &  &  &  &  &  & \makecell{1.46*\\[-3pt]{\footnotesize$+0.31$}}\\
\quad Data \& Research &  &  &  &  &  &  &  &  &  &  & \\
\textbf{Years of experience} (per 5 years) &  &  &  &  &  &  &  &  &  &  & \\
\textbf{Task interdependence} (per point, 1--5) & \makecell{1.13*\\[-3pt]{\footnotesize$+0.06$}} &  &  &  &  &  &  &  &  &  & \\
\textbf{Working with AI} (per point, 1--5) &  & \makecell{1.12*\\[-3pt]{\footnotesize$+0.05$}} &  &  &  &  &  &  &  &  & \\
\midrule\multicolumn{12}{@{}l}{\textbf{\textit{Psychographics:}}}\\[0pt]
\textbf{Function} (baseline: judges and evaluates) &  &  &  &  &  &  &  &  &  &  & \\
\quad Makes their own rules &  &  &  &  &  &  &  &  &  &  & \\
\quad Works to the rules given &  &  &  & \makecell{1.24*\\[-3pt]{\footnotesize$+0.14$}} & \makecell{1.24*\\[-3pt]{\footnotesize$+0.18$}} & \makecell{1.29*\\[-3pt]{\footnotesize$+0.18$}} &  &  & \makecell{1.22*\\[-3pt]{\footnotesize$+0.19$}} & \makecell{1.22*\\[-3pt]{\footnotesize$+0.19$}} & \\
\textbf{Form of sequencing} (baseline: ranks, prioritizes) &  &  &  &  &  &  &  &  &  &  & \\
\quad One thing at a time &  &  &  &  &  &  &  &  &  &  & \\
\quad Several at once, unranked &  &  &  &  &  &  &  &  &  &  & \\
\quad No fixed order &  &  &  &  &  &  &  &  &  &  & \\
\textbf{Focus-orientation} (baseline: whole picture) &  &  &  &  &  &  &  &  &  &  & \\
\quad The detail &  &  &  &  &  &  & \makecell{1.15*\\[-3pt]{\footnotesize$+0.09$}} &  &  &  & \\
\textbf{Openness to new ways} (baseline: new ways) &  &  &  &  &  &  &  &  &  &  & \\
\quad Established ways &  & \makecell{0.76**\\[-3pt]{\footnotesize$-0.13$}} &  &  &  &  &  &  &  &  & \\
\midrule\multicolumn{12}{@{}l}{\textbf{\textit{Stances on who counts as a co-worker and what they are for:}}}\\[0pt]
\textbf{Weight on getting ahead} (per 10 of 100 points) &  &  &  &  & \makecell{0.88***\\[-3pt]{\footnotesize$-0.11$}} & \makecell{0.94*\\[-3pt]{\footnotesize$-0.05$}} &  &  & \makecell{0.91***\\[-3pt]{\footnotesize$-0.09$}} & \makecell{0.93***\\[-3pt]{\footnotesize$-0.07$}} & \makecell{0.94*\\[-3pt]{\footnotesize$-0.06$}}\\
\textbf{AI feels like a co-worker} (per point, 1--5) & \makecell{1.09*\\[-3pt]{\footnotesize$+0.05$}} &  &  & \makecell{1.23***\\[-3pt]{\footnotesize$+0.13$}} & \makecell{1.13**\\[-3pt]{\footnotesize$+0.11$}} & \makecell{1.13*\\[-3pt]{\footnotesize$+0.09$}} & \makecell{1.07*\\[-3pt]{\footnotesize$+0.04$}} &  & \makecell{1.15**\\[-3pt]{\footnotesize$+0.13$}} & \makecell{1.15***\\[-3pt]{\footnotesize$+0.13$}} & \makecell{1.15**\\[-3pt]{\footnotesize$+0.12$}}\\
\midrule Pseudo-$R^{2}$ & 0.010 & 0.012 & 0.013 & 0.025 & 0.024 & 0.020 & 0.008 & 0.012 & 0.029 & 0.022 & 0.023\\
\bottomrule
\end{tabular*}
\Description{A regression-results table for the AI-referent sample of 802 respondents. The leftmost column lists predictors; eleven narrow columns with vertically rotated headings report separate fractional logit models for Craftsperson, Safe Pair of Hands, Fair Dealer, Explorer, Stand-Up Colleague, Perseverer, Navigator, Straight Talker, Glue, Satisficer, and Room-Maker, in that order. The first two rows give each model's baseline odds and corresponding implied probability. Horizontal rules separate three predictor sections. Work context includes discipline, years of experience, task interdependence, and working with AI. Psychographics includes rule orientation, sequencing, focus orientation, and openness to new ways, with alternative styles indented beneath each dimension. The final section, Stances on who counts as a co-worker and what they are for, weights getting ahead and how strongly AI feels like a co-worker. Reference categories and units of change appear beside predictor labels. Each populated predictor cell shows an odds ratio with significance stars above an implied difference on the original one-to-five rating scale. Blank cells indicate results that did not survive the Benjamini--Hochberg correction. Numeric predictors are centered on AI-referent means. A final row reports pseudo-R-squared for each model.}
\end{table*}

\textbf{Overall priorities:} The baseline worker rated \arch{The Explorer} highest for a human co-worker (82\%), followed by \arch{The Stand-Up Colleague} and \arch{The Perseverer} (80\% each), and \arch{The Satisficer} lowest (62\%). For AI, they rated \arch{The Safe Pair of Hands} highest (88\%), followed by \arch{The Craftsperson} and \arch{The Straight Talker} (84\% each), and \arch{The Glue} and \arch{The Satisficer} lowest (57\% and 60\%).

Against that reference, 77 of 385 contrasts survived correction (49 human-side and 28 AI-side), with at least one in 21 of the 22 models. Human-side differences were broad: discipline, interdependence, help from AI, and attention to detail each raised priority across several archetypes. AI-side differences were fewer and concentrated in relational archetypes, especially \arch{The Glue}, \arch{The Satisficer}, \arch{The Room-Maker}, and \arch{The Stand-Up Colleague}. We unpack these differences below, categorized by knowledge-worker characteristics.

% End of slimdown section

\subsection{Work context}
\label{sec:whowantswhat:variation}

\textbf{Discipline.} Compared to software engineers, go-to-market and customer-facing respondents (sales, customer service, consulting, and marketing) placed higher priority on human co-workers who were honest, reliable, and fair-minded (\arch{The Fair Dealer}: $+0.26$ points), who persevered (\arch{The Perseverer}: $+0.20$), made pragmatic decisions (\arch{The Satisficer}: $+0.17$), learned through curiosity and collaboration (\arch{The Explorer}: $+0.15$), and mapped strategy and progress, timed and scoped information, and repaired drift (\arch{The Navigator}: $+0.15$).

IT and infrastructure respondents also placed higher priority on \arch{The Fair Dealer} ($+0.22$) and \arch{The Perseverer} ($+0.21$), as well as co-workers who connected and cared for people (\arch{The Glue}: $+0.24$), were available, knowledgeable, and candid (\arch{The Straight Talker}: $+0.22$), and owned the outcome and shared credit (\arch{The Stand-Up Colleague}: $+0.21$).

Business and operations respondents (business and program administration, finance, and legal) also prioritized \arch{The Fair Dealer} ($+0.29$) and \arch{The Stand-Up Colleague} ($+0.17$), and co-workers who anticipated needs, organized and paced the work (\arch{The Room-Maker}: $+0.25$), and whose work was thorough, transparent, and open to challenge (\arch{The Craftsperson}: $+0.16$). P105-H, in business operations, tied fair dealing to holding a line under pressure: \q{In the Deal Desk we often have to reject deal structures that elicit emotional reactions from Sales Teams and Customers. If we can't hold up under pressure, it drives inconsistent outcomes as we scale concessions across customers.} No human-side archetype contrasts for product \& design or data \& research survived correction.

For AI, only three disciplines had archetype-level contrasts with the baseline worker. Technical and infrastructure respondents prioritized AI that anticipated needs, organized the work, and shared the conversational turn (\arch{The Room-Maker}: $+0.32$), as did business and operations respondents ($+0.31$), who also placed higher priority on \arch{The Fair Dealer} ($+0.21$).

% Other disciplines were comparable to the baseline in terms of their priority of co-worker archetypes. At the quality level, however, product and design respondents placed higher priority on \Ab{Takes genuine ownership} ($+0.45$) and lower priority on \In{Transparent about reasoning} ($-0.43$). P666-H noted that transparency \q{feels unimportant as long as co-workers own accountability for their work.}
Go-to-market and customer-facing respondents prioritized AI that connected with and cared for people (\arch{The Glue}: $+0.28$), rating its \Be{Social skill}, \Be{Recognizes individuality}, \Be{Empathy}, and \Be{Caring} higher. In their free-text answers, respondents connected these qualities to individualized interactions and relationships at work. P98-A, in a customer-facing role, noted: \q{There is always a pivot from customer to customer in my role. Custom responses are an essential part of the picture}. They also prioritized AI that anticipated their needs (\arch{The Room-Maker}: $+0.22$), and one that made pragmatic decisions (\arch{The Satisficer}: $+0.20$).

\textbf{Years of experience.} Each five additional years in the industry lowered the priority respondents placed on a human \arch{Navigator} ($-0.04$ points). No AI-side archetype contrast survived correction; at the quality level, more experienced respondents rated an AI's \Be{Equal conversational turn-taking} higher.

\textbf{Task interdependence.} The more a respondent's work depended on others, the more they prioritized eight human-side archetypes spanning learning and candid expertise (\arch{The Explorer}: $+0.13$; \arch{The Straight Talker}: $+0.12$), thorough work, perseverance, and follow-through (\arch{The Craftsperson}: $+0.10$; \arch{The Perseverer}: $+0.10$; \arch{The Safe Pair of Hands}: $+0.09$), and ownership, fair dealing, and coordination (\arch{The Stand-Up Colleague}: $+0.09$; \arch{The Fair Dealer}: $+0.07$; \arch{The Navigator}: $+0.07$). Their open-ended responses identified facilitation and coordination as what made interdependent work possible. For AI, by contrast, only \arch{The Craftsperson} rose ($+0.06$): as work became more interdependent, respondents placed more priority on AI being transparent, adapting to change, working thoroughly, and challenging their thinking.

\textbf{Working with AI}. The more AI helped a respondent get work done, the more they prioritized an AI that was sound in judgment and followed through (\arch{The Safe Pair of Hands}: $+0.05$). For a human co-worker, their priority rose on archetypes spanning coordination and pacing (\arch{The Navigator}: $+0.09$; \arch{The Room-Maker}: $+0.09$), and ownership and care (\arch{The Stand-Up Colleague}: $+0.07$; \arch{The Glue}: $+0.07$), with \Co{Facilitation} being the quality they rated highest in particular. P82-H, whom AI helps a lot, noted: \q{AI has made it cheap to produce plausible-looking work, so the bottleneck has shifted from generating to verifying. The co-workers I rely on most are the ones who catch the wrong assumptions before it ships}.

\subsection{Psychographics}

We modeled four dimensions of Sternberg's thinking-style inventory as independent dispositions toward organizing work (recall \S\ref{sec:survey}): \emph{Function}, whether someone prefers to make their own rules, work to the rules they are given, or judge and evaluate based on context; \emph{Form of sequencing}, whether they work on one thing at a time, rank and prioritize, run several things at once, or work in no fixed order; \emph{Focus-orientation}, whether they attend to the detail or the whole picture; and \emph{Openness to new ways}, whether they prefer new or established ways of working~\citep{sternberg1997thinking}.

\textbf{Function.} Relative to respondents who judge and evaluate (baseline), those who preferred making their own rules placed lower priority on four human-side archetypes spanning pacing, shared understanding, and context awareness (\arch{The Room-Maker}: $-0.21$; \arch{The Safe Pair of Hands}: $-0.16$) and legibility and fair dealing (\arch{The Navigator}: $-0.15$; \arch{The Fair Dealer}: $-0.14$). P282-H, who preferred to make their own rules, noted: \q{I don't care if someone's work is observable or not. I don't need to know someone is doing something, so long as it gets done at the end of the day.}

For AI, respondents who preferred to work to the rules they were given placed higher priority on five archetypes involving care, patience, affect management, consensus building, and pragmatic decisions (\arch{The Glue}: $+0.19$; \arch{The Satisficer}: $+0.19$), ownership and perseverance (\arch{The Stand-Up Colleague}: $+0.18$; \arch{The Perseverer}: $+0.18$), and learning through collaboration (\arch{The Explorer}: $+0.14$). Their free-text answers explained a boundary around authority: those who made their own rules more often raised concerns about losing control or treating AI as a person, whereas those who worked to given rules emphasized coordination and adaptation to change while keeping the final decision with the person. P582-A, who preferred to work to given rules, wanted AI to \q{collaborate with people to solve problems, build and maintain a shared understanding across stakeholders, and help navigate disagreements constructively}.

\textbf{Form of sequencing} produced only two contrasts, both for human co-workers. Compared to respondents who preferred ranking and prioritizing tasks, those who preferred doing one thing at a time prioritized co-workers who were available, had domain expertise,  resisted reflexive agreement (\arch{The Straight Talker}: $+0.27$), and decided when work was good enough and sought help when needed (\arch{The Satisficer}: $+0.22$).

\textbf{Focus-orientation.} Compared with respondents who preferred the whole picture, detail-oriented respondents prioritized seven human co-worker archetypes more highly. These differences concerned judgment about when to stop and visibility into progress, uncertainty, and provenance (\arch{The Satisficer}: $+0.18$; \arch{The Navigator}: $+0.14$); confirming understanding, inviting input, and recovering from error (\arch{The Safe Pair of Hands}, \arch{The Craftsperson}, and \arch{The Fair Dealer}: each $+0.13$); and ownership, shared credit, and openness to challenge (\arch{The Stand-Up Colleague}: $+0.13$; \arch{The Perseverer}: $+0.12$). For AI, the difference remained only for \arch{The Navigator} ($+0.09$); detail-oriented respondents placed greater priority on visibility into its progress, provenance, uncertainty, and limits.

\textbf{Openness to new ways.} Compared with respondents who preferred new ways, those who preferred established ways prioritized human co-workers who made pragmatic decisions, recognized when work was good enough, and sought help when needed (\arch{The Satisficer}: $+0.14$); P135-H wanted co-workers to \q{know when the outcome has been achieved and stop fine-tuning or wasting brain cells over minuscule details}. They also placed lower priority on \arch{The Stand-Up Colleague} ($-0.11$), whose qualities included managing conflict and not blaming for honest mistakes. P506-H held that \q{If a mistake was made, blame should be assigned. It's what is done afterwards to ensure understanding and future prevention}. For AI, they placed lower priority on sound judgment and follow-through (\arch{The Safe Pair of Hands}: $-0.13$) because they kept both for themselves; P330-A: \q{my POV is that AI should support the work and minimize the menial tasks that need to be done but should NOT own decisions (or make them for you)}. 
% Sternberg describes this style as a preference for existing procedures and minimal change \citep{sternberg1997thinking}.

\subsection{Stances on who counts as a co-worker and what they are for}

\textbf{Pictured their manager or the people they manage as co-workers.} Some respondents read ``co-worker'' to include their manager or their reports, so we modeled whom they pictured as a control: 98\% pictured peers, 42\% also pictured their manager, and 16\% pictured people they manage. Those who included their manager brought managerial expectations of direction, ownership, care, and follow-through into the co-worker referent, prioritizing \arch{The Straight Talker} ($+0.14$); \arch{The Stand-Up Colleague}, \arch{The Navigator}, and \arch{The Room-Maker} ($+0.13$ each); \arch{The Glue} ($+0.12$); and \arch{The Safe Pair of Hands} ($+0.11$). No contrast for respondents who pictured people they manage survived correction.

\textbf{Weight on getting ahead over getting along.} Recall that respondents had divided 100 points between the two motives of working life: getting ahead and getting along (\S\ref{sec:surveydesign}). For every 10-point shift toward getting ahead, they placed lower priority on five AI-side archetypes spanning shared credit and accountability (\arch{The Stand-Up Colleague}: $-0.11$), care and facilitation (\arch{The Glue}: $-0.09$), pragmatic decisions (\arch{The Satisficer}: $-0.07$), pacing and participation (\arch{The Room-Maker}: $-0.06$), and ownership and perseverance (\arch{The Perseverer}: $-0.05$). On the human side, \Be{Personable} was the only quality rated lower; no human-side contrast survived correction.

\textbf{AI felt like a co-worker.} The more AI had felt like a co-worker, the more respondents prioritized AI-side archetypes spanning exploration, care, and pragmatic decisions (\arch{The Explorer}, \arch{The Glue}, and \arch{The Satisficer}: $+0.13$ each); participation, accountability, and perseverance (\arch{The Room-Maker}: $+0.12$; \arch{The Stand-Up Colleague}: $+0.11$; \arch{The Perseverer}: $+0.09$); and goal-oriented, thorough, and inspectable work (\arch{The Craftsperson}: $+0.05$; \arch{The Navigator}: $+0.04$). Interestingly, the relational archetypes that fell as weight on getting ahead increased; all rose as AI felt more like a co-worker. P286-A, for whom AI had felt very much like a co-worker, noted: \q{I ask AI to recognize my individuality, and respond in a manner I speak to make it feel more like a real coworker as I interact with it on different topics.}

\vspace{1mm}
\begin{takeaway}
\tkw{RQ3 --- Variation} Work context raised human-side expectations broadly, whereas AI-side differences concentrated in relational work and dependable execution. Among thinking styles, the clearest differences were that respondents who made their own rules wanted less process from people, those who worked to given rules wanted more of a teammate from AI, and detail-oriented respondents set a higher bar for co-workers overall. Weight on getting ahead lowered the relational AI archetypes, and feeling that AI is a co-worker raised them; picturing a manager raised expectations of direction, ownership, care, and follow-through.
\end{takeaway}
\vspace{2mm}
\section{RQ4 --- Work Etiquette: What do knowledge workers expect from co-workers who share AI-supported work?}
\label{sec:emerging}

\begin{table*}[t]
\footnotesize
\setlength{\tabcolsep}{7pt}
\renewcommand{\arraystretch}{1.15}
\caption{Taxonomy of AI work etiquette capturing what recipients expect of co-workers who share AI-supported work. Four themes organize 15 sub-themes: the pass workers owe before work goes out under their name, the declaration they owe alongside it, the cost they pass on when they skip their duty, and the standing they need before using AI at all. In the theme and sub-theme labels, ``you'' refers to the worker producing and sharing the work.}
\Description{A three-column table showing the taxonomy of AI-work etiquette with headings Theme, Sub-theme, and Description of the codes it consolidates. Four sections, separated by horizontal rules, organize 15 sub-themes. Italicized theme labels occupy the left column, sub-themes the middle column, and explanations the right column. In the labels, ``you'' refers to the worker producing and sharing AI-supported work. The first theme concerns the pass workers owe before AI-supported work goes out under their name. Its five sub-themes cover reading and verifying the work; being able to judge and answer for it; pushing back on the model; remaining accountable for checks that fit the stakes; and retaining the worker's own voice. The second theme concerns the declarations workers owe alongside the work. Its three sub-themes cover disclosing AI involvement and its extent, reporting what was checked and showing the work, and stating the cognitive investment the work deserves. The third theme concerns the cost workers pass on when they skip their duty. Its three sub-themes describe how the sender's saving becomes the reader's bill, how recipients search for evidence of care and human review, and why ``please verify'' does not discharge the sender's obligations. The fourth theme concerns the standing workers need before using AI at all. Its four sub-themes cover supplying the work's context, using AI as a bounded amplifier within the worker's competence, protecting confidential and private information and using approved tools, and avoiding automation of relationships or using AI on co-workers without their knowledge.}
\label{tab:etiquette}
\begin{tabular}{@{}
  >{\raggedright\arraybackslash}p{0.15\textwidth}
  >{\raggedright\arraybackslash}p{0.32\textwidth}
  >{\raggedright\arraybackslash}p{0.47\textwidth}@{}}
\toprule
\textbf{Theme} & \textbf{Sub-theme} & \textbf{Description of the codes it consolidates}\\
\midrule
\textit{The pass you owe before sharing AI-supported work}
  & \textbf{You read and verified it before you sent it}
  & Read or edit the work, and check its claims against sources, data, or a running build, before it changes hands. Hold it to the standard you would meet unaided.\\[7pt]
  & \textbf{You can judge it and can answer for it}
  & Understand the work well enough to explain and defend it, use AI only where you could have evaluated the result unaided, and keep exercising this judgment.\\[7pt]
  & \textbf{You pushed back on the model rather than blindly accepting it}
  & Interrogate the output instead of forwarding it, and do not carry the model's authority into a disagreement with someone who knows the domain better.\\[7pt]
  & \textbf{You are accountable for the check, and the check fits the stakes}
  & Keep an answerable person in the deciding seat, treat a further model pass as no discharge of that duty, and scale the depth of checking to what the work can break.\\[7pt]
  & \textbf{It still sounds like you}
  & Return the work to your own voice and register, which for some respondents means leaving no detectable trace of the tool at all.\\
\midrule
\textit{The declaration you owe alongside it}
  & \textbf{Say the tool was involved, and how much}
  & Disclose that AI was used and what share of the work it did, and do not claim its output as your own or let an agent act unattended in your name.\\[7pt]
  & \textbf{Say what you checked and show the working}
  & Report which sources were used, which parts were verified and to what depth, and what remains unresolved.\\[7pt]
  & \textbf{State the cognitive investment the work deserves}
  & State the effort behind the artefact and label unfinished work; the same declaration extends to the person assigning work, who states the investment they expect.\\
\midrule
\textit{The cost you pass on when you skip your duty}
  & \textbf{Your saving is the reader's bill}
  & Do not move skipped checking, unnecessary length, or unprocessed output onto the recipient; select, compress, or interpret what you send.\\[7pt]
  & \textbf{The forensics of care}
  & Undeclared work leaves the recipient foraging the surface for evidence that a person reviewed it: stock phrasing reads as neglect, and a confident surface reads as concealment, ultimately damaging trust.\\[7pt]
  & \textbf{``Please verify'' is not a disclosure}
  & A label reporting unspent attention transfers the duty in writing, and the judgment that follows is passed on the sender rather than the artefact.\\
\midrule
\textit{The standing you need before using it at all}
  & \textbf{The context is your work, not the AI's}
  & Fit the work to its project, customer, and audience, and supply the prompts, grounding, and domain knowledge that get it there.\\[7pt]
  & \textbf{Use AI as a bounded amplifier}
  & Use AI to clarify communication or extend the reach of a person's contribution, while remaining within the bounds of what they can competently judge.\\[7pt]
  & \textbf{Protect what is confidential, private or IP-related, and use the approved tools}
  & Keep customer data, personal information, and protected material out of the tool, and stay inside approved systems and organizational policy.\\[7pt]
  & \textbf{Do not automate the relationship or aim the AI at co-workers}
  & Do not delegate interpersonal communication or use the AI on co-workers to profile, rank, or ingest them without their knowledge.\\
\bottomrule
\end{tabular}
\end{table*}
\vspace{3mm}

The interviews surfaced obligations between co-workers when preparing and sharing AI-supported work that the BACI framework did not capture. We therefore included an open-ended question in the surveys asking: \textit{``What do you consider good vs. bad professional practice when a coworker uses AI on work that reaches you?''} 
% We received 1,018 responses to this question, of which 991 were codeable.

Recall, we used the AI-council-supported reflexive thematic analysis procedure described in \S\ref{sec:interviews}. Two authors approved the codebook before the AI council independently coded every response; assignments required agreement from at least two analysts. Across the codes, Krippendorff's $\alpha$ ranged from 0.83 to 1.00, with a median of 0.94, and 95\% of assignments survived to consensus. This analysis produced 15 sub-themes across 4 themes, summarized in Table~\ref{tab:etiquette} and detailed below.

\subsection{The pass you owe before sharing AI-supported work}

\vspace{1mm}
\textbf{You read and verified it before you sent it.} 630 respondents (63.6\%) treated reading and verification as the minimum owed before a handoff. Reading meant making a pass over the material; verification meant checking claims against facts, figures, sources, and data, or running and testing technical artefacts. P491-H objected bluntly: \q{I find it rude when coworkers chuck dirty slop at me without reviewing the work first. If they didn't take the time to read it, why should I?} Participants applied the same requirement to customer communication, criticizing \q{Taking a [AI tool] answer, not checking it, and pasting it directly into an email to a customer is sloppy and lazy.}

\vspace{1mm}
\textbf{You can judge it yourself and can answer for it.} For 216 respondents (21.8\%), a pass was insufficient unless the sender understood the work well enough to explain it and remain accountable for it. Some considered AI use legitimate only when the sender could judge the result independently; others worried that habitual reliance displaced the sender's own thinking and learning. P299-H drew a clear line: \q{Vet what you share. If you don't understand it, don't share it.} That responsibility did not end when the work changed hands. P720-H rejected AI as an excuse when questions arose: \q{If there are parts I don't understand, I expect my coworker to be able to explain it instead of saying, "I dunno, [AI] generated it."} P127-A described a comic exchange: \q{a co-worker used AI to summarize a doc I wrote, and then asked me if the summary is correct :) I view that as bad practice!}

\vspace{1mm}
\textbf{You pushed back on the model rather than blindly accepting it.} 165 respondents (16.6\%) objected when a co-worker treated the model's answer as settling the matter and were unwilling to challenge or overrule it. P311-H criticized \q{trusting [AI] is always right and not validating or challenging what it spit out is against work hygiene.} This deference became an interpersonal violation when someone used the model's authority to dismiss the recipient's domain expertise. P504-H objected to \q{asking AI to do legal analysis and then arguing with me based on that, in my area of expertise, and presuming AI has weighed the analysis properly, is insulting.}

\vspace{1mm}
\textbf{You are accountable for the check, and the check fits the stakes.} For 87 respondents (8.8\%), saying that AI output should be checked left two questions open: who should check it, and how much checking was enough. Some reserved sign-off for a person accountable for the result. A second AI pass did not necessarily count as review, particularly when it added nothing the recipient could not do themselves. P561-A criticized \q{when a coworker becomes an AI wrapper themselves. Sometimes folks don't review my code and just do a generic AI review which I could have done myself.} The required scrutiny also rose with the stakes involved. P41-A explained: \q{it depends on context \& the amount of review done with AI. The lesser the stakes, the more I'm ok with it.}

\vspace{1mm}

\textbf{It still sounds like you.} For 59 respondents (6.0\%), AI-supported work needed to retain the sender's voice as evidence of engagement. P771-A explained: \q{I still need to feel like a person is involved, and so when there is no tuning of the language/structure/content, it's obvious. My subconscious picks it up at this point and stops trusting it.} Some took this expectation even further, arguing that good AI use should leave no trace of AI involvement.

\vspace{3mm}
\subsection{The declaration you owe alongside it}

\vspace{1mm}
\textbf{Say the tool was involved, and how much.} For 118 respondents (11.9\%), disclosure meant stating both that AI was used and how much it contributed. This helped recipients calibrate scrutiny and kept AI output from passing as the sender's alone. P631-A treated it as ordinary source attribution: \q{If a colleague used AI, I would expect them to share that, just as I would expect them to say, "well I googled the answer."} However, disclosure was not an alternative for engagement with the work. P654-A noted: \q{With AI being verbose and chatty, a good practice is to list it as co-authored and then really read through it; if they don't understand it, I won't either}.

\vspace{1mm}
\textbf{Say what you checked and show the working.} For 48 respondents (4.8\%), naming AI was insufficient; senders also had to identify which sources and claims they had checked, what they knew to be true, and what remained unresolved. P164-H drew this boundary explicitly: \q{I don't mind if someone uses AI to work through something, but if you are sharing it, be clear what \& how much you validated.} Without that context, recipients had to uncover the unchecked work themselves. P475-H described receiving \q{Broken links, unvetted assumptions, and more} when co-workers neither reviewed AI output nor explained how much they had reviewed it.

\vspace{1mm}
\textbf{State the cognitive investment the work deserves.} Disclosure also needed to convey how much thought the work had received and how much of it was expected from the recipient. For 17 survey respondents (1.7\%), rough AI output was acceptable when its draft status, purpose, and quality bar were explicit. P675-H described \q{a simple note saying "this is a vibe-coded prototype, what do you think? I'll validate XYZ for the next iteration" sets clear expectations about the investment needed}. The interviews surfaced a reciprocal obligation. P10 wanted requesters to specify how much cognitive effort the task warranted: \q{for example, this is how much cognitive load I expect you to carry with this, or, I'm thinking this is a 6 to 8 hour exercise. Use AI as a wireframe, but I expect this to be a heavy cognitive lift for you}.

Interestingly, these disclosure expectations were not universal. 50 respondents (5.0\%) rejected the premise of special disclosure or reversed the obligation. For some, AI use had become too ordinary to warrant explanation. Others treated failure to use AI, or to share effective ways of using it, as a professional lapse. P346-A argued: \q{In this day and age I think it's irresponsible to not use AI if you have access to it.} Some shifted responsibility toward employers that required, measured, or rewarded AI use. P318-A acknowledged attribution as good practice while describing organizational pressure that made nondisclosure routine: \q{you can tell everyone just dumps their work into [AI] (I do too, because they tell me to.) Everyone's tech designs went from babblespeak with poor grammar to babblespeak with better grammar. Every sentence reads like: "It’s not just a tool — it’s a paradigm shift."}

\subsection{The cost you pass on when you skip duty}

\vspace{1mm}
\textbf{Your saving is the reader's bill.} For 187 respondents (18.9\%), AI did not eliminate work so much as move it downstream. When senders skipped review or editing, recipients had to verify claims, cut through unnecessary volume, or redo work they could have asked AI to perform themselves. P703-H described this imbalance: \q{It's bad professional practice and quite disrespectful in my opinion when a coworker makes me do more work understanding what they're sending than they spent creating it.} Merely returning an AI answer to the recipient's question created the same problem. P103-A warned: \q{Be honest, and assume I've already asked the AI. Just [AI]ing my question back to me is a terrible practice. It's as insulting as the old "let me google that for you" link from the 2000s.}

\vspace{1mm}
\textbf{The forensics of care.} When senders did not disclose how they had handled AI output, recipients searched the artefact for evidence of human attention, losing trust in the sender when they found none. Among 149 respondents (15.0\%), 73 called out unreviewed output slop, while 76 treated stock formatting and phrasing as \emph{visible AI tells} that nobody reviewed. P287-H singled out one familiar tell: \q{Leaving the LONG a** hyphens in a communication. It reads as fake and inauthentic and sometimes makes me distrust the information.} Some also noted that polish and fluency could conceal the same omitted review. P795-A identified: \q{a bad practice is forwarding a polished draft they haven't verified, where the confident tone hides that nobody has checked the specifics, that moves the work of catching errors onto me, without the warning that it's needed.} 
% P724-A further distrusted fluency that exceeded the sender's apparent ability: \q{I believe that the finished product looks good; however, sometimes it far outdoes the intelligence of the sender, therefore falsely reflecting their expertise, which makes me trust their genuineness}.

\vspace{1mm}
\textbf{``Please verify'' is not a disclosure.} For 14 respondents (1.4\%), attaching a mere warning to unreviewed output merely turned disclosure into an instruction for the recipient to complete the omitted check. P160-H rejected this subcontracting: \q{I HATE when a coworker sends me AI-generated content with the flag "ai-generated content, please verify for accuracy" disclaimer in the email or created doc - it's as if they're subcontracting the work to ME! If you don't have time to proofread the content before sending, then why should I take the time when I receive it?} Such repeated incidents changed how later work, and its sender, were judged. P391-H explained: \q{If I can tell it's slop, I can't trust it's correct and I lose respect/trust in that person. It's hard to repair that as well, and I'll be on guard for any future work from that person}.

\subsection{The standing you need before using it at all}

\vspace{1mm}
\textbf{The context is your work, not the AI's.} 96 respondents (9.7\%) treated fit to the project, customer, and audience as the sender's contribution. They also emphasized that the worker was responsible for prompt craft, grounding, and guardrails, and noted that generic or off-topic output was therefore a failure of the handoff. P27-H clarified that this responsibility was not simply about possessing the most knowledge, but about communicating the right information to the relevant audience: \q{Before AI, the best resources in the company weren't necessarily those who knew the most, but those who could communicate the right information effectively to different audiences.}

\vspace{1mm}
\textbf{Use AI as a bounded amplifier.} 48 respondents (4.8\%) identified two ways AI could legitimately amplify a co-worker’s contribution: helping them communicate more clearly and enabling them to contribute more broadly. In both cases, the sender needed sufficient context and competence to judge the output. P209-A welcomed AI \q{when my coworker uses [it] to surface themes or details from complicated conversations that happened in email, recorded calls, etc.} Participants similarly valued AI when it \textit{expanded horizontal skills}, but warned that such \q{contributions could fall below standard when the sender lacked domain knowledge to judge quality, as in vibe coding} (P413-A).

\vspace{1mm}
\textbf{Protect what is confidential, private or IP-related, and use the approved tools.} 32 respondents (3.2\%) required co-workers to protect confidential, personal, customer, and proprietary information and to use only approved AI tools. P363-H  paired institutional permission with personal responsibility: \q{use the systems as laid out by company standards and thoroughly check the outputs produced by it no different than what you'd do for a new college grad}.

\vspace{1mm}
\textbf{Do not automate the relationship or aim the AI at co-workers.} 58 respondents (5.9\%) objected to delegating personal communication or using AI to profile, rank, or evaluate co-workers without their knowledge. Participants noted that \q{a 1-1 message or retirement note carried the sender's attention as part of the message; delegating it to AI could therefore feel distancing or signal that the receiver was not worth the effort} (P132-H).
At its limit, automation removed both people from the exchange. P552-H feared \q{AI writes your message and will also essentially get a response from another AI; real communication is going out the door. It just shows me that I'm not important enough for this person to actually write an email themselves.} 

\vspace{1mm}
\begin{takeaway}
\tkw{RQ4 --- Work etiquette} Respondents held senders to reading and verifying the work, answering for it, and disclosing what the tool did and what was checked, and what remained unfinished. Otherwise, the sender's saved effort became the recipient's verification work, and these handoffs reduced trust in later work. Respondents accepted AI as a way to extend capability or improve communication, but required protected information to stay within approved tools and resisted automating relationship-bearing work.
\end{takeaway}

\section{Discussion}
\label{sec:discussion}

Our study found that workers carried many co-worker expectations across humans and AI, yet assigned them different priorities and obligations, which further varied with workers' characteristics (\S\ref{sec:framework}--\ref{sec:whowantswhat}). Based on these, we discuss implications for building worker-centric AI, adapting AI to workers' diverse contexts, and extending these lessons to mixed human--AI teams and multi-agent systems.

Our study further surfaced obligations around responsible co-working as a first-class concern of AI-native work (\S\ref{sec:emerging}). Accordingly, we detail considerations for workplace design around which human contributions organizations should preserve and how workers, processes, and teams can share responsibility.

\subsection{Implications for AI tool design}
\label{sec:aidesign}

\subsubsection{\textbf{Building worker-centric AI}}

An extensive line of work argues that AI support should be evaluated against human-centric objectives: whether it builds workers' skills, preserves their agency, supports collaboration, and sustains meaningful work~\citep{buccinca2024towards,choudhuri2026ai,shneiderman2020humancentered}. This argument draws on work design research linking workers' autonomy and competence to their motivation and performance~\citep{hackman1976motivation}, and labor economics arguing that AI must extend workers' capabilities to deliver gains beyond ``labor arbitrage''~\citep{brynjolfsson2023turing}.

Our findings could help translate these objectives into concrete qualities against which AI tools could be designed and evaluated. For example, the archetypes participants prioritized for AI (Figure~\ref{fig:quadrant}; \S\ref{sec:prevalence}) point to several such qualities: a \arch{Safe Pair of Hands} that maintains a shared understanding of goals, constraints, and evolving context and makes informed decisions aligned with them; a \arch{Craftsperson} that produces thorough, adaptable work, follows safe and secure practices, and is transparent; a \arch{Fair Dealer} that listens actively, avoids unfair bias, regulates its behavior, and recovers from errors; and a \arch{Straight Talker} that resists sycophantic agreement and points to relevant expertise when needed.

Participants connected these qualities to their own position in the work, i.e., being able to judge AI's contributions, keep the work moving, and avoid repeated context-setting or rework that could negate AI's promised benefits. Worker-centricity may also benefit from a counterintuitive division of labor, i.e., AI absorbing more of the cognitive overhead of rote coordination (e.g., maintaining context, detecting misalignment, and recovering from error) so that workers can exercise agency where it matters, in setting direction and making consequential judgments.

\subsubsection{\textbf{Adapting to workers' diverse context}}

Workers' differing expectations around what they value in their co-workers reinforce the need for AI support to adapt to worker diversity (\S\ref{sec:whowantswhat}). Existing tools already provide ways to shape their behavior. For example, AI tools support saved project instructions and standing approvals that auto-approve actions in a session~\citep{anthropicPersonalization,anthropicPermissions}. Learning from a user's behavior is indeed an established HAI principle~\citep{amershi2019guidelines}. 

Our findings give this adaptation a more specific focus. The qualities participants prioritized in a co-worker varied across several layers of their context, differing around how much AI challenges them, how much process and visibility it provides, how much it coordinates, whether and when it offers benevolence, and how far it may act independently. Designs should account for this diversity to help workers articulate how they want an AI to contribute, and to guide its behavior accordingly. 
Doing so requires accounting for several layers of user context (recall \S\ref{sec:whowantswhat}): 

\begin{enumerate}[leftmargin=*, itemsep=2pt, topsep=3pt]
  \item \textit{Stance toward AI and collaboration}: participants for whom AI had felt like a co-worker asked more of it relationally, while those weighing accomplishment over relationships asked less.
  \item \textit{Work context and current scenario}: what participants wanted differed by their discipline, their experience, and how much they used AI, and also by the purpose of the exchange.
  \item \textit{Thinking and working style}: detail-oriented participants wanted more visibility into AI's progress, sources, uncertainty, and limits, and those who worked to given rules wanted more of a ``teammate'' from AI.
\end{enumerate}

Organizational conditions bound all three, since workplace policy governs which tools may be used, which data is protected, and what may be inferred about a person.

This understanding should be developed with the worker and remain open to correction. Yet, workers may not want to specify all of it, so an AI should infer a starting point from these layers. Following research on scrutable user modeling~\citep{kay2013scrutable}, that inference could be made visible and correctable: for instance, a worker might revise ``You prefer to work independently'' to ``On this project, I want regular check-ins and help coordinating with co-workers.'' Correction of this kind may also support reflection and planning~\citep{kay2019personal}, and workers' metacognitive flexibility~\citep{tankelevitch2024metacognitive}, by letting them reconsider how to approach a task and what help to seek from AI.

\subsubsection{\textbf{From human--AI to team--AI collaboration}}

AI support for one worker can create work for another. Respondents frequently described receiving poorly prepared AI-supported work that they then had to check and repair. These accounts suggest that AI should also consider the needs of the people who receive the work alongside those of the immediate user. For example, in a shared ecosystem, when helping someone prepare a document for a colleague, an AI could use context that the colleague has chosen to share about their role, knowledge, and intended use to identify what evidence and explanation the document needs.

Supporting several people, however, requires keeping their priorities distinct. In a shared tool, one person's latest interaction should not silently override the team's goals or agreements. If co-workers need different levels of detail or disagree about scope, the tool should make those differences explicit and help the people involved decide how to proceed. Such support also requires control over how each person is represented: workers should be able to inspect and correct what the tool assumes about them and choose what others can see.

Furthermore, participants described progressing farther on their own with AI, sometimes drifting from the team's shared understanding and deliberately having to check in with colleagues to restore it (recall \S\ref{sec:framework:cooperativeness}). A team tool should thus support these check-ins by flagging when work appears to depart from agreed goals or rely on assumptions the team has not discussed. This would support the repair of common ground \citep{klein2005common,johnson2014coactive}, while leaving the discussion and decisions to the people involved.

Disagreement, in this context, deserves particular attention. Respondents gave low priority to conflict management, consensus building, facilitation, finding win-wins, turn-taking, and affect management. Participants also anticipated that, as AI becomes more agreeable, ordinary disagreement from a colleague might feel rude. 

This raises a fundamental question around \textit{whether working with an agreeable AI routinely changes workers' willingness to engage with human disagreement, or the value they place on support for doing so}. For design, this suggests that a team tool should make competing assumptions and evidence available for discussion and avoid presenting agreement as settled before the people involved have reached it.

\subsubsection{\textbf{From team--AI to AI--AI collaboration}}

A further possibility in AI-native workplaces is that one worker's AI exchanges work with a colleague's or team's AI. Our study did not observe such exchanges, but the qualities workers valued suggest several design hypotheses. Assistants could be expected to retain relevant context, follow through on commitments, report sources and uncertainty, identify conflicting assumptions, and stay within the authority people have given them.

These exchanges should also preserve differences between the people represented. If one AI requests a final recommendation while another provides an exploratory draft, they should identify that mismatch before treating the work as ready for use. Decisions beyond their delegated authority should likewise return to the people responsible.

Relational qualities would matter substantively through the effects of these exchanges on people's work and relationships: whether colleagues' priorities are represented accurately, whether unresolved differences remain visible, and whether people retain control over commitments made on their behalf. Future work should test whether these designs improve coordination and help workers meet their obligations to one another.

\subsection{Implications for AI-native work design}\label{sec:workdesign}

Where AI can perform knowledge work, organizations still face a choice about which contributions to preserve and develop in people. Participants' expectations point to situated taste and discernment, willingness to disagree, ownership and accountability, emotional attentiveness, and care for co-workers (recall \S\ref{sec:ranking}).

These expectations suggest that hiring and team formation should consider how people assess and improve work, explain their choices, and respond to questions, alongside what they produce. A work sample could, for example, ask a candidate to critique an AI-supported draft and identify which checks require further expertise.

Training could similarly focus on developing these capabilities through practice in framing tasks, checking sources, questioning AI suggestions, recognizing the limits of one's expertise, and correcting errors. Workers also need opportunities to develop ideas and learn with co-workers. Participants described AI tools taking over tasks previously assigned to junior co-workers, raising concerns about how those co-workers would learn. Where this occurs, organizations could preserve opportunities for supported practice and mentoring so that junior workers can develop the judgment they will later be expected to exercise.

Lastly, responsible handoffs should be a part of how work is organized. Participants objected when a co-worker's saved effort became their own checking, interpretation, or repair, and they interpreted omitted preparation as a lack of care. Rather than relying on individual conscientiousness, organizations can build this etiquette into practice at three levels: 
% the worker's own, what AI and surrounding processes can carry, and what teams and organizations require.

\begin{enumerate}
\item \emph{\textbf{Workers' practices.}} Read and shape work before sending it, check it to the agreed standard, and identify what remains unfinished or uncertain. Fit the contribution to its audience and retain the personal attention the relationship calls for. Question the AI's suggestions and remain able to explain the resulting work. When judging its adequacy requires expertise the sender lacks, arrange appropriate review and make that need explicit.

\item \emph{\textbf{Support from tools and processes.}} Keep evidence and assumptions available, help workers check claims and identify gaps, and prepare work for its recipient. An expandable handoff summary could accompany the artefact, identifying who is accountable, what AI contributed, what was checked and by whom, unresolved questions, and the action requested. The AI could draft this summary from available work history, with the sender confirming and completing it. In particular, the summary should distinguish exploratory work, material awaiting expert review, and work ready for its intended use.

\item \emph{\textbf{Team and organizational agreements.}} Agree who checks what, what counts as adequate review, and who is responsible for addressing problems afterward. Build these agreements into review and approval practices, with time and access to appropriate expertise. Set expectations for permitted tools, protected information, and tasks requiring personal engagement. Expectations should be explicit in both directions: senders identify AI's contribution and completed checks, while requesters state where they expect cognitive investment or personal attention. Respondents disagreed about when disclosure was needed, and prior work similarly identifies social penalties for disclosing AI use \citep{schilke2025transparency,reif2025penalty}. Teams therefore need shared disclosure norms and organizational support to address penalties for following them.
\end{enumerate}

% Responsible handoffs make the recipient's needs part of how work is prepared, reviewed, and shared. They also make clear that the benefits of AI-supported work depend on whether it helps the team, including those who must interpret, check, and act on the result.

Overall, responsible handoffs can help translate workers' expectations of AI and of one another into shared practices for preparing, checking, and passing on work. In an AI-native workplace, doing work well includes giving the next person what they need to understand it, judge its limits, and use it responsibly.

\subsection{Limitations}
\label{sec:limitations}

% As with every empirical study, ours has limitations.
\vspace{1mm}
\textbf{Construct validity.} The phrase ``great co-worker'' could mean different things to different people. We intentionally left it undefined in an effort to capture participants' perceptions; there is no universal definition, and our results aggregate the views of people with diverse experiences. We also report what knowledge workers said they prioritize, not how they behave toward a co-worker who has or lacks a quality. We treat this as motivation for future work: the priorities and obligations participants articulated provide testable criteria. Our survey items were derived from the interviews and grounded in established theory, so respondents rated qualities that knowledge workers themselves had named (\S\ref{sec:framework}). Still, surveys can introduce bias or misunderstanding. We addressed this risk by involving practitioners in design, piloting the survey in two rounds, randomizing blocks, adding attention checks, and screening implausibly short and incomplete responses (recall \S\ref{sec:analysis}). 

\vspace{1mm}
\textbf{Internal validity.} Our goal was not to capture a purely random sample of interviewees; we chose a stratified approach to capture a wide range of responses from a diverse group, varying discipline, industry experience, and AI use (recall \S\ref{sec:interviews}). Within the strata, the interview participants were self-selected; even so, all 22 interviewees belonged to different teams. It is impossible to cover the entire variety of teams that work on \site's products and services, but the fact that the interviewees came from different teams gives us confidence that we covered as much ground as possible, given pragmatic restrictions of access. We then used the interviews to inform our survey, which was deployed broadly enough to provide representativeness and a sample large enough for statistical inference. Self-selection bias is possible, since those with stronger views may be more likely to participate. Moreover, unmeasured differences between the two groups of participants may have contributed to referent contrasts. We strengthened validity by member-checking the framework with interviewees, and triangulating quantitative results with qualitative data and theory where applicable. As with all empirical work, results reflect self-reported perceptions.

% Since we built one regression model per archetype and referent, we controlled the false discovery rate across the contrasts~\cite{benjamini1995controlling}. There is ``no universally accepted approach for dealing with the problem of multiple comparisons''~\cite{mcdonald2014handbook}, and any correction trades false discoveries against missed ones. Ours errs towards the latter, so modest differences may not have survived. The effects reported in \S\ref{sec:whowantswhat} are those that did, and we report the per-quality differences by respondent context in the supplemental~\cite{supplemental}.

\vspace{1mm}
\textbf{AI-assisted analysis.} Using AI models in qualitative analysis sits in tension with human interpretation, theoretical sensitivity, and judgment~\cite{braun2006using,braun2019reflecting}. Our pipeline does not attempt to replace interpretive qualitative work. Rather, it separates pattern discovery across a large corpus, which the models supported, from interpretive judgment about what those patterns mean, which we retained; models served as discovery and coding instruments under human supervision (recall safeguards reported in \S\ref{sec:analysis}). Further, we limited the risk of automation bias by ordering the work: two authors open-coded the transcripts first, and the council applied the author-approved codebook afterward to check for missed qualities. Every assignment was reviewed by two authors before it took effect, and approximately 95\% of the authors’ original assignments remained. 

Our practices offer a practical way to use AI in large-scale qualitative analysis while preserving human control over interpretation. However, they do not claim to resolve whether AI can interpret human meaning~\cite{schroeder2025large}, and we kept that boundary explicit throughout the analysis. Verbatim quotes throughout the paper allow readers to assess whether themes are grounded in participants' responses. 

\vspace{1mm}
\textbf{External validity.} Our analysis comes wholly from one organization. Although multi-organization studies are valuable, access to employees makes such studies difficult, and responses may not be as candid with ``outsiders''. We therefore do not claim to represent all knowledge workers. Instead, we provide an in-depth account of a large organizational context. Single-case studies, like ours, have advanced scientific discovery~\cite{flyvbjerg2004five} and produced insights in the social sciences and HAI research~\cite{kalliamvakou2019manager,choudhuri2026ai}. \site{} employs tens of thousands of knowledge workers across varied work contexts, and our stratified random sampling improves generalizability. Still, our findings likely transfer better to large technology organizations than to smaller or regulated ones. They also reflect the tools, policies, and practices present during data collection. Future work should test transferability in other contexts, and to encourage replication, we have made our study instruments available~\cite{supplemental}.

\section{Conclusion}
\label{sec:conclusion}

Knowledge workers wanted the same core from humans and AI: craft, a shared understanding of the work and its goals, collaborative problem-solving, and reliable, thorough work open to challenge. Beyond that core, expectations diverged. Of AI, workers wanted ready-made expertise, non-sycophancy, adaptability, and observable, redirectable actions and reasoning. From people, they expected discernment, candid disagreement, ownership and accountability, emotional attentiveness, and reciprocal care. These priorities further varied with workers' context, styles, and relationship to AI.

The AI-native workplace also brought new expectations for responsible co-working. Workers expected co-workers to read and verify what they passed on, challenge the AI's output, and stand behind the resulting work. They also expected them to disclose how much the AI contributed and what had been checked, so that effort saved upstream did not become a cost borne downstream.

Our findings raise fundamental questions about how to design an AI-native workplace, and how working with AI changes what it means for any of us to be a great co-worker. As a first step, we suggest designing AI tools around the qualities workers seek and the differences among those workers, preserving the judgment, accountability, and care still expected of people, and making the handoff of AI-supported work a shared responsibility. Great co-workers make others' work possible. That's still the whole test.

\begin{acks}
We thank all participants for their time and insights. Rudrajit Choudhuri and Sam Yu-Te Lee were Microsoft interns during this research.
\end{acks}

\bibliographystyle{ACM-Reference-Format}
\bibliography{references}

% \ifshowsupp
% %% Everything from here to \fi is supplemental: long, generated, and not needed to follow
% %% the argument. Set \showsuppfalse in the preamble to leave it out.

% \clearpage\onecolumn
% \appendix

% \section{The interview protocol}
% \label{app:protocol}
% \input{supplemental/appendix_protocol}

% \section{The survey instrument}
% \label{app:instrument}
% \input{supplemental/appendix_instrument}

% \section{The 75 great co-worker qualities}
% \label{app:battery}
% \input{tables/appendix_battery}

% \section{Instrument-validation diagnostics}
% \input{tables/tab_validation}

% \section{The thinking and working style inventory, as fielded}
% \label{app:thinkingstyles}
% \input{supplemental/supp_thinking_styles}

% \section{Which qualities each kind of worker emphasizes}
% \label{app:effects}
% \input{supplemental/supp_quality_effects}

% \section{Why each quality was rated high or low}
% \label{app:reasons}
% \input{supplemental/supp_quality_reasons}

% \fi

\end{document}